\documentclass[11pt,a4paper]{article}

\usepackage{jheppub}
\usepackage{graphicx}

\usepackage{xurl} 
\usepackage{hyperref} 
\usepackage{bm}
\usepackage{siunitx}
\usepackage{array}
\newcolumntype{d}{S[table-format=1.4(4)]}\usepackage{amsmath}
\usepackage{amssymb}
\usepackage{needspace}
\usepackage{color}
\usepackage{changes}
\usepackage{float}
\usepackage{subfig}
\usepackage{multirow}
\usepackage{algorithm}
\usepackage{algorithmic}
\usepackage{tabularx}
\usepackage[section]{placeins}  
\usepackage{booktabs}

\newcommand{\el}{\vspace*{0.3cm}}

\title{Stochastic Path Sampler For Lattice Field Theory}

\author[a,1]{Shiyang Chen,}
\emailAdd{2222895@swansea.ac.uk}
\affiliation[a]{Centre for Quantum Fields and Gravity, Department of Physics, Swansea University, SA2 8PP, Swansea, United Kingdom}

\author[b,c,1]{Moxian Qian,}
\emailAdd{mqian@students.uni-mainz.de}
\affiliation[b]{School of Science and Engineering, The Chinese University of Hong Kong, Shenzhen (CUHK\-Shenzhen), Guangdong, 518172, China}
\affiliation[c]{Johannes Gutenberg University Mainz, 55128 Mainz, Germany}
\author[a]{Gert Aarts,}
\emailAdd{g.aarts@swansea.ac.uk}
\note{These authors contributed equally to this work.}
\author[d]{Biagio Lucini,}
\emailAdd{b.lucini@qmul.ac.uk}
\affiliation[d]{School of Mathematical Sciences, Queen Mary University of London, Mile End Road, London, E1 4NS, UK}

\author[b,e]{Kai Zhou}
\emailAdd{zhoukai@cuhk.edu.cn}
\affiliation[e]{School of Artificial Intelligence, The Chinese University of Hong Kong, Shenzhen (CUHK-Shen\-zhen), Guangdong, 518172, China}

\date{\today}

\abstract{ In lattice field theory, target distributions are known only up to normalization, $\tilde{\pi}(\phi)\propto e^{-S(\phi)}$, while the partition function is generally intractable. Markov chain Monte Carlo simulations often become inefficient near phase transitions or the continuum limit due to critical slowing down. In this work, we propose a novel sampler based on nonequilibrium thermodynamics, called Stochastic Path Sampler (SPS), which can generate configurations for the unnormalized target distribution and without a need for training data.  The central idea of the SPS is to establish a trajectory-level balance for learnable forward and backward stochastic dynamics between two equilibrium states—the prior and target distributions. This is achieved by minimizing the path-space variational free energy, equivalently an entropy-production upper bound, defined by the log-ratio of forward and auxiliary backward trajectory measures, to enhance the reversibility of the forward and backward processes. The learned forward process provides independent proposals, which are subsequently corrected by an extended-space Independence Metropolis--Hastings step. 
In two-dimensional $\phi^4$ theory, we demonstrate that our neural sampler can achieve the same sampling quality as HMC but with a much shorter autocorrelation time in the pseudocritical region. This sampler offers a novel stochastic-quantization-inspired route to data-free proposal construction for lattice field theory
by leveraging a variational free-energy principle derived from path-space irreversibility.
}

\begin{document}

\maketitle

\section{Introduction}
Lattice field theory is a core tool for studying strong interactions, critical phenomena, and quantum many-body systems. Traditional Markov chain methods such as Hybrid Monte Carlo (HMC) typically face two classical challenges:
(i) \emph{critical slowing down} causes a dramatic increase in autocorrelation time;
(ii) \emph{topological freezing} prevents the Markov chain from adequately exploring different topological or symmetry sectors within finite time.

In recent years, as the computational power of GPUs has become increasingly affordable and various generative models have been proposed, it has become possible to use neural networks to address critical slowing down and topological freezing for an unnormalized target distribution $\tilde{\pi}^\ast(\phi)=e^{-S(\phi)}$, where $S(\phi)$ is the action of the target distribution. Currently, generative models adopted in lattice field theory (LFT) can be divided into two categories based on their requirements for training reference data: variational free‑energy based training models and data‑driven training models, see, e.g., the reviews \cite{Cranmer:2023xbe,Aarts:2025gyp,Tomiya:2025quf}.

Variational free‑energy based training models include normalizing flows (NFs) \cite{rezende2015variational}, continuous normalizing flows (CNFs) \cite{chen2018neural} and autoregressive networks~\cite{Wu:2019elz,Wang:2020hji}. NFs have been used to generate lattice field configurations in scalar \cite{Albergo_2019,Nicoli:2020njz}, fermionic \cite{albergo2021flow} and gauge theories \cite{Kanwar_2020,Boyda_2021,Foreman:2021ljl,Albergo_2022}.  Related variational autoregressive approaches have also been explored for scalar $\phi^4$ theory \cite{Zhou:2018ill,Qian:2025VANphi4}. CNFs are based on a single neural ODE layer with specially designed basis functions  \cite{Gerdes_2023, dehaan2021scalingmachinelearningquantum}, yielding a higher effective sample size (ESS) in $\phi^4$ theory.  However, when the target distribution has nontrivial topological structure (such as multiple rings, multimodal peaks, or multiple topological sectors), the connected manifold output by the network struggles to cover the entire support set, easily leading to mode collapse \cite{Chen:2025kfo,hackett2025flowbasedsamplingmultimodalextendedmode}. Therefore, a gauge equivariant neural network architecture or a special kernel basis needs to be introduced to address topological issues.  Relatedly, Fourier-parameterized flow models have been investigated to better respect periodic structure and alleviate multimodal sampling difficulties in path-integral settings, e.g., by introducing Fourier transforms into the flow to generate Feynman paths under periodic conditions \cite{Chen:2022ytr,Zhou:2023pti}. Stochastic normalizing flows \cite{wu2020stochastic,Caselle_2022} include Langevin diffusion steps into NFs, to address critical slowing down \cite{caselle2022stochasticnormalizingflowslattice, caselle2025stochasticnormalizingflowseffective,Caselle_2025,Bulgarelli:2024brv}.
 
Data‑driven training models mainly consist of generative adversarial networks (GAN) \cite{Zhou:2018ill} and diffusion/score‑based models (DMs) \cite{song2019generative}. GANs usually suffer from training instability since the adversarial training over two competing neural networks. DMs are based on Langevin dynamics and supervised training. In LFT, they have been used  in $\phi^4$ \cite{Wang:2023exq}, $U(1)$ gauge theory \cite{Zhu:2025pmw,Vega:2025hgz}, non-abelian gauge theories \cite{Aarts:2026zzr,Alharazin:2026lcb,Komijani:2026lan},  and also for data generated from complex-valued distributions \cite{Aarts:2025lpi}.  These methods often rely on labeled samples for supervised training and can serve as amortized samplers (training is expensive, but sampling is cheap) in physical applications, trained on reference samples in a supervised manner. Recent work projects a trained flow-matching velocity onto lattice-operator and Fourier-mode bases to identify magnetization and low-momentum directions, guiding a normalized, invertible proposal that requires no neural-network evaluation during sampling \cite{Qian:2026OperatorSpectroscopy}.

In the broader machine-learning literature, there are several related approaches that use data-free diffusion-based samplers, minimizing a path-space Kullback-Leibler objective.  Continual Repeated Annealed Flow Transport Monte Carlo (CRAFT) \cite{pmlr-v162-matthews22a} combines learned invertible flow maps with sequential Monte Carlo along a prescribed sequence of annealed target distributions. Non-equilibrium transport samplers (NETS) \cite{Albergo:2024trn} introduce a learned transport drift on top of Langevin dynamics along an explicitly prescribed time-dependent annealing potential. Other examples are the  Path Integral Sampler \cite{zhang2022pathintegralsamplerstochastic},  Denoising Diffusion Samplers  \cite{vargas2023denoisingdiffusionsamplers},  optimal-control formulations of diffusion-based generative modeling \cite{berner2024optimalcontrolperspectivediffusionbased,richter2024improvedsamplinglearneddiffusions}, and Controlled Monte Carlo Diffusions \cite{vargas2025transportmeetsvariationalinference}. In contrast, SPS learns both the forward and  backward stochastic dynamics directly between a simple Gaussian prior and the target distribution, while the intermediate marginal distributions are generated implicitly by the learned dynamics rather than specified through a sequence of intermediate equilibrium targets. By minimizing the forward--backward path-space KL divergence, SPS adaptively searches for transport trajectories with enhanced reversibility, rather than following a prescribed annealing path. The present work can therefore be viewed as a stochastic-quantization-inspired adaptation of this broader family of path-space variational samplers to lattice field theory, with exactness restored by a trajectory-level Independence Metropolis--Hastings correction.

The Clausius inequality is one of the fundamental formulations of the second law of thermodynamics. It states that irreversible processes are accompanied by nonnegative entropy production, while equality is reached only in the reversible limit. Beyond its original thermodynamic context,  that principle has become a general framework for characterizing irreversibility, dissipation, and the directionality of nonequilibrium processes. Its applications range from materials science \cite{ColemanNoll1963}, ecological systems \cite{Kleidon2010,Dyke2010}, biological physics and neuroscience \cite{Skinner2021,Karbowski2024}, economics and urban systems \cite{Jakimowicz2020,Purvis2019}, to information theory   \cite{Landauer1961,SagawaUeda2008,Parrondo2015}. For sampling problems, the Clausius inequality admits a particularly useful path-space interpretation. The entropy production can be viewed as a measure of the discrepancy between forward and backward trajectory distributions. Minimizing this discrepancy therefore provides a variational route to suppress irreversibility and to construct stochastic dynamics that transports a simple prior distribution toward a target Boltzmann distribution more efficiently.

In this paper, we propose the Stochastic Path Sampler (SPS), a neural sampler inspired by nonequilibrium thermodynamics and stochastic quantization \cite{Parisi:1980ys,Damgaard:1987rr}.
Unlike supervised diffusion models, SPS does not require reference configurations from HMC; it uses only evaluations of the target action $S(\phi)$. This is achieved by realizing that the $\log$ ratio of the forward and backward path measures over the entire trajectory can be viewed as an entropy. Then by reducing the entropy, the irreversibility between the forward and backward processes is decreased, so that the forward and backward processes admit a global balance on the entire trajectory, implying that the distribution sampled by the forward process follows the target distribution. The paper is organised as follows. In Section~\ref{sec:2}, we define the path measure via the discretized Langevin equation, and then introduce the loss function of SPS and the Independence Metropolis--Hastings algorithm. In Section~\ref{sec:3}, we apply our network model to the $\phi^4$ model, compute observables, and compare them with the results from HMC. Section~\ref{sec:4} is the conclusion. Some technical details are given in Appendix \ref{app:duality_relation}. Appendix \ref{sec:Network} contains details of the network architecture, sampling and training employed. Appendix \ref{app:C} presents magnetization histograms used to assess the mode coverage of the uncorrected SPS proposals near the pseudocritical region. Finally, Appendix~\ref{app:ess_generation_depth} discusses the dependence of the effective sample size and the acceptance rate on the generation depth and training progress.

\section{Nonequilibrium Thermodynamics on Path Space}
\label{sec:2}

In statistical physics and field theory, we are often faced with the task of generating samples 
representative of a complicated target distribution (such as the Boltzmann distribution $\pi^*=\tilde{\pi}^\ast/Z$ with the normalization factor $Z$, referred to as state B) from a known initial distribution (such as a Gaussian distribution $\pi_0$, referred to as state A). These states are typically associated with forward and backward processes driven by appropriate dynamics and the process connecting the two equilibrium states is generally nonequilibrium in nature.

\label{sec:thermo}
\begin{figure}[t]
  \centering
  \includegraphics[width=0.8\linewidth]{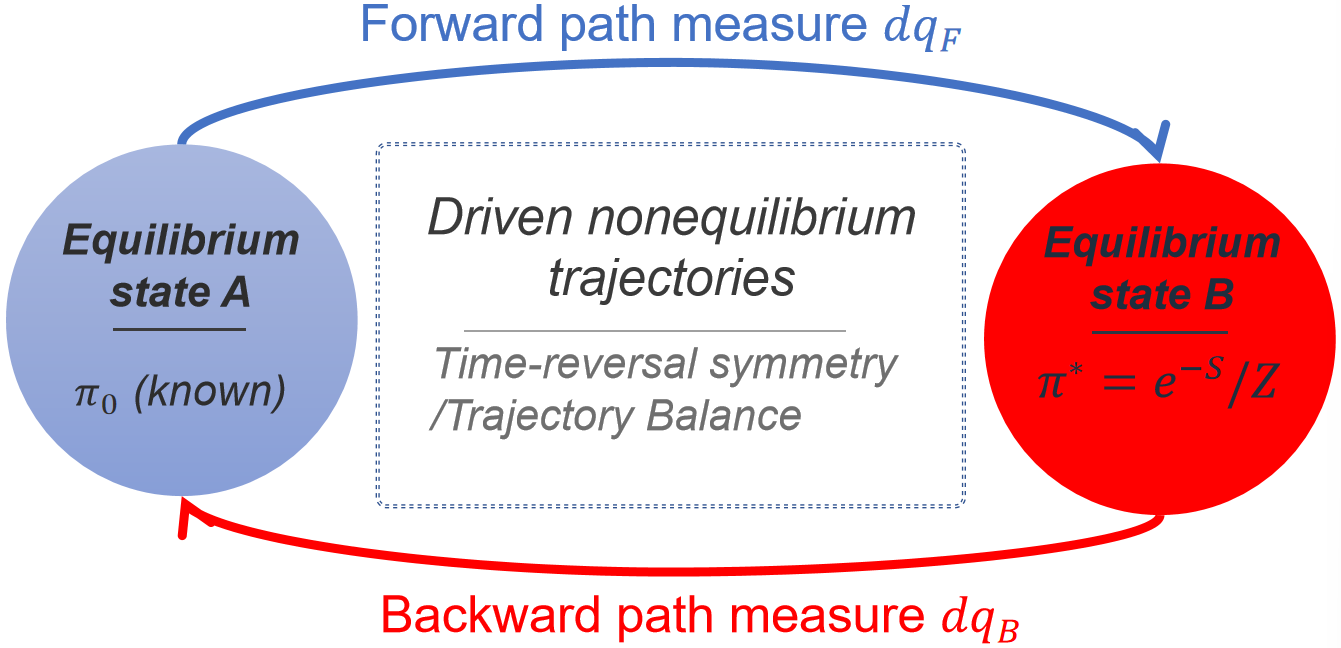}
  \caption{\textbf{ Equilibrium endpoints and nonequilibrium path measures.} A driven forward process transports an initial equilibrium distribution $\pi_0$ (state A) to a target equilibrium $\pi^\ast$ (state B), inducing a forward path measure $dq_{\mathrm F}$ over trajectories. A backward dynamics starting from $\pi^\ast$ induces $dq_{\mathrm B}$ on the same trajectory space.}
  \label{fig:equilibrium_noneq_schematic}
\end{figure}
As illustrated in Fig.~\ref{fig:equilibrium_noneq_schematic}, we define two path measures: the forward process transports $\pi_0$ to $\pi^*$, inducing a path distribution $dq_{\text{F}}$; conversely, the backward process attempts to revert from $\pi^*$ back to $\pi_0$, inducing a path distribution $dq_{\text{B}}$.

The core objective of this work is to achieve reversibility under the forward and backward path measures. Specifically, our goal is to establish time-reversal symmetry (Trajectory Level Balance \cite{bengio2021gflownet,bengio2023gfnfoundations, malkin2022trajectory}). This means we need to adjust the dynamics to minimize the irreversibility between the forward and backward processes, forcing the ratio of the two path measures to approach a constant (corresponding to the free energy difference) over the entire trajectory space. By achieving this global trajectory balance, we can ensure that the samples generated by the forward process strictly follow $\pi^*$.

\subsection{Entropy Production on Path Space}
The Clausius inequality provides a mathematical formulation of the Second Law of Thermodynamics. It states that, for a cyclic transformation, the integral of the infinitesimal heat exchanged with the environment divided by temperature is always nonnegative, and vanishes only for a reversible cyclic process \cite{Pachter2024}. This inequality establishes the thermodynamic notion that irreversibility can be quantified by a nonnegative entropy-production term. In the present work, we adopt its stochastic-thermodynamic counterpart, namely entropy production on path space.

For a stochastic process evolving between microscopic configurations, entropy production measures the asymmetry between the forward trajectory ensemble and its corresponding reversed ensemble. This provides a stochastic-thermodynamic formulation of nonequilibrium irreversibility, without requiring the construction to be based on a macroscopic heat-exchange process.

In stochastic thermodynamics, transitions between microscopic states are characterized by a local thermodynamic affinity $\delta A_\tau(t)$. Accordingly, the stochastic entropy production of the system is given by the ensemble average of the accumulated affinity, together with a boundary term $\log (\pi_0/\pi^\ast)$,
\begin{equation}
\Sigma
=
\left< \int_0^T dt \,\delta A_\tau(t) \right>
+
\left< \log \frac{\pi_0}{\pi^\ast} \right> .
\label{eq:ensemble_average}
\end{equation}
Here $\Sigma \ge 0$ characterizes the irreversibility of the path-space dynamics. Equality corresponds to a reversible path ensemble, for which the forward and reversed trajectory probabilities are time-reversal symmetric.

\subsection{Discrete Langevin Paths and Learnable Forward/Backward Kernels}

We use $\bm{s}$ to denote a sample; in our setting $\bm{s}$ coincides with the configuration $\phi$. We write $\bm{s}_i$ for the state of the sample at step $i$ along a trajectory and $\bm{s}(t)$ for the corresponding time-indexed sample. We work on the normalized time interval $[0,1]$, uniformly discretized into $T$ steps. The discrete time points are defined as $t_i = i \Delta t$  for  $i = 0, 1, \dots, T$, where $\Delta t = 1/T$.

A discrete forward Langevin update from step $i$ to $i+1$ is written as
\begin{equation}
\begin{aligned}
\bm{s}_{i+1}
\equiv \bm{s}_i +
 \sigma^2_{\theta}(t_{i})\bm{K}_{\theta,\mathrm{F}}(\bm{s}_{i},t_{i}) \Delta t
&+ \sigma_\mathrm{\theta}(t_{i})\,\bm{\xi}_i\sqrt{\Delta t},
\end{aligned}
\label{eq:discrete_langevin_forward}
\end{equation}
where $\bm{\xi}_i \sim \mathcal{N}(\bm{0}, \bm{I})$. 
For simplicity of analysis, we assume that the diffusion coefficient $\sigma_\mathrm{\theta}(t_{i})$  is a scalar and depends only on time:
\begin{equation}
\sigma_{\theta,\mathrm{F}}(t_i) = \sigma_{\theta,\mathrm{B}}(t_{i+1})\equiv \sigma_{\theta}(t_i).
\end{equation}
While the diffusion coefficient is usually chosen manually in Langevin-based samplers \cite{Wang:2023exq}, here we parameterize it by a neural network and optimize it jointly with the drift terms. Quantities determined by a neural network carry the subscript $\theta$, representing 
the neural net parameters.

The corresponding forward transition kernel is denoted as $q_\mathrm{F}(\bm{s}_{i+1} | \bm{s}_i)$. In the Metropolis-Adjusted Langevin Algorithm (MALA) \cite{roberts1998optimal,girolami2011riemann} and in stochastic quantisation \cite{Parisi:1980ys,Damgaard:1987rr}, the drift term $\bm{K}_{\theta,\mathrm{F}}(\bm{s},t_{i})$ is taken as the gradient of the action,
\begin{equation}
    \bm{K}_{\theta,\mathrm{F}}(\bm{s},t_{i}) =-\nabla_{\bm{s}} S(\bm{s}),
\end{equation}
but in our current setting, it is approximated by a neural network with time embedding. 

Given a prior $\pi_0$ (Gaussian prior in this work), the discrete Langevin update (\ref{eq:discrete_langevin_forward}) induces a stochastic trajectory,
\begin{equation}
\tau: \bm{s}_0 \to \bm{s}_1 \to \dots \to \bm{s}_{T},
\end{equation}
with the forward path distribution
\begin{equation}
dq_{\mathrm{F}}(\tau)
= \pi_0(\bm{s}_0)\prod_{i=0}^{T-1}
q_{\mathrm{F}}(\bm{s}_{i+1}|\bm{s}_i)\, d\tau,
\label{eq:forward_path_prob}
\end{equation}
where $q_{\mathrm{F}}(\bm{s}_{i+1}|\bm{s}_i)$ is the Gaussian kernel defined by Eq.~\eqref{eq:discrete_langevin_forward}.

For the same path $\tau$, we similarly define a discrete backward process with a learnable transition kernel $q_\mathrm{B}(\bm{s}_i | \bm{s}_{i+1})$ which pulls the target distribution back to the prior as 
\begin{equation}
\begin{aligned}
\bm{s}_{i}
\equiv \bm{s}_{i+1} + \sigma_{\theta}^2 (t_{i})&\bm{K}_{\theta,\mathrm{B}}(\bm{s}_{i+1},t_{i})\Delta t + \sigma_{\theta}(t_{i})\bm{\eta}_{i+1}\sqrt{\Delta t}.
\end{aligned}
\label{eq:discrete_langevin_backward}
\end{equation}
with learnable backward drift term $\bm{K}_{\theta,\mathrm{B}}(\bm{s}_{i+1},t_{i})$ via another neural net and $\bm{\eta}_{i+1} \sim \mathcal{N}(\bm{0}, \bm{I})$. The corresponding backward path distribution is
\begin{equation}
dq_{\mathrm{B}}(\tau)
= \pi^\ast(\bm{s}_T) \prod_{i=0}^{T-1}
q_{\mathrm{B}}(\bm{s}_{i}|\bm{s}_{i+1})\, d\tau.
\label{eq:backward_path_prob}
\end{equation}

\subsection{Path Probability Ratio as Discrete Entropy  Production}
The stepwise affinity for the nonequilibrium system is defined as the log ratio of forward and backward transition amplitudes,
\begin{equation}
   \delta A(t_i)= \log \frac{ q_\mathrm{F}(\bm{s}_{i+1}|\bm{s}_{i})}{ q_\mathrm{B}(\bm{s}_{i}|\bm{s}_{i+1})},
\label{eq:Work_Production}
\end{equation}
and the accumulated affinity with the boundary term (after time discretisation) is defined as
\begin{equation}
    \Sigma(\tau) = \sum_{i=0}^{T-1} \log \frac{ q_\mathrm{F}(\bm{s}_{i+1}|\bm{s}_{i})}{ q_\mathrm{B}(\bm{s}_{i}|\bm{s}_{i+1})} + \log\frac{\pi_0(\bm{s}_0)}{\pi^\ast(\bm{s}_T)}.
    \label{eq:Entropy_Production}
\end{equation}
This demonstrates irreversibility of the nonequilibrium system; the larger $\Sigma(\tau)$, the stronger the forward--backward inconsistency, and hence the more irreversible the process.

A natural objective to reduce irreversibility is therefore the Clausius entropy,
\begin{equation}
\mathbb{E}_{d\, q_{\mathrm F}}[\Sigma(\tau)] = \int dq_{\mathrm{F}}(\tau)\log \frac{dq_{\mathrm{F}}(\tau)}{dq_{\mathrm{B}}(\tau)},
\label{eq:loss}
\end{equation}
which is the Kullback-Leibler (KL) divergence \cite{kullback1951information} on path space. The KL divergence can be optimized by adjusting the learnable drifts $\bm{K}_{\theta,\mathrm{F/B}}$ and  the diffusion coefficient $\sigma_\theta$. When the normalization factor for $\pi^\ast$ is intractable, we adopt the unnormalized distribution $\tilde{\pi}^*$ and the unnormalized backward path measure, and use the KL divergence between the forward and unnormalized backward path measure,
\begin{equation}
    D_{KL} = \int dq_{\mathrm{F}}(\tau)\log \frac{dq_{\mathrm{F}}(\tau)}{d\, \tilde{q}_{\mathrm{B}}(\tau)}\, \quad \mbox{with} \quad d \tilde{q}_{\mathrm{B}}(\tau)
= \tilde{\pi}^\ast(\bm{s}_T) \prod_{i=0}^{T-1}
q_{\mathrm{B}}(\bm{s}_{i}|\bm{s}_{i+1})\, d\tau,
\label{eq:kl_unnormalized}
\end{equation}
to reduce the irreversibility.

Minimizing $\mathbb{E}_{q_{\mathrm F}}[\Sigma(\tau)]$ is equivalent to reducing irreversibility and improving sampling efficiency, which can be achieved by learning the drifts in the forward and backward dynamics. Via the Jarzynski equality \cite{Jarzynski:2013qyb,Pachter2024,Caselle:2016wsw}, under perfect training, the KL divergence admits an entropy-free energy decomposition,
\begin{equation}
\mathbb{E}_{q_{\mathrm F}} \!\left[ \, \sum_{i=0}^{T-1} \log \frac{ q_\mathrm{F}(\bm{s}_{i+1}|\bm{s}_{i})}{ q_\mathrm{B}(\bm{s}_{i}|\bm{s}_{i+1})} + \log\frac{\pi_0(\bm{s}_0)}{\tilde{\pi}^\ast(\bm{s}_T)}\right]= - \log Z.
\label{eq:JE}
\end{equation}
Therefore, by minimizing Eq.~\eqref{eq:loss}, the forward path distribution and the backward path distribution become closer, constraining the ratio between forward and backward trajectory probabilities to match a single global constant, which corresponds to the log normalizer $\log Z$ of  \replaced{$\tilde{\pi}^\ast$}{$\pi^\ast$}. This leads to
\begin{equation}
\frac{\prod_{i=0}^{T-1} q_{\mathrm{F}}(\bm{s}_{i+1}\mid \bm{s}_i)\,\pi_0(\bm{s}_0)}{\prod_{i=0}^{T-1} q_{\mathrm{B}}(\bm{s}_{i}\mid \bm{s}_{i+1})}
\propto \tilde{\pi}^\ast(\bm{s}_T).
\label{eq:tb_ratio_main}
\end{equation}
Note that in the continuous limit, Eq.~\eqref{eq:tb_ratio_main} is equivalent to a line integral, 
\begin{equation}
\int d\bm{s}(t)\cdot
\big(\bm{K}_{\theta,\mathrm{B}}(\bm{s},t)
+ \bm{K}_{\theta,\mathrm{F}}(\bm{s},t)\big) + \log\pi_0(\bm{s}_0)
\approx \log\tilde{\pi}^\ast(\bm{s}_T) + \text{const}.
\label{eq:line_integral_main}
\end{equation}
This expression shows that the learned stochastic dynamics constructs a path-dependent transport map from the prior to the target distribution. When the irreversibility vanishes, the accumulated drift contribution compensates for the difference between the prior density and the unnormalized target density, leaving only the global constant $-\log Z$. In general, however, finite model capacity, finite training time, and discretization errors prevent exact trajectory balance, so that the variational objective remains above its ideal value \(-\log Z\). For a trajectory \(\tau\), we define the path-dependent endpoint estimator
\begin{equation}
q_\theta(\mathbf{s}_T;\tau)
\equiv
\frac{
\prod_{i=0}^{T-1}
q_{\mathrm{F}}(\mathbf{s}_{i+1}\mid \mathbf{s}_i)
}{\prod_{i=0}^{T-1}
q_{\mathrm{B}}(\mathbf{s}_{i}\mid \mathbf{s}_{i+1})
}\pi_0(\mathbf{s}_0).
\label{eq:tb_ratio_general}
\end{equation}
At finite training, \( q_\theta(\mathbf{s}_T;\tau)\) generally depends on the full trajectory \(\tau\), not only on the endpoint \(\mathbf{s}_T\). This path dependence reflects the residual irreversibility between the forward and backward path measures. In the ideal limit of exact trajectory balance, the ratio of forward and backward path probabilities becomes a global constant over path space. Consequently, the dependence on the particular trajectory \(\tau\) disappears, and \( q_\theta(\phi;\tau)\) reduces to an endpoint proposal density \(q_\theta(\phi)\), which is proportional to the unnormalized target density \(\tilde{\pi}^\ast(\phi)\).

This continuous-time form is closely related to Nelson's duality relation, which connects the forward and backward drifts of a diffusion process through the score of the intermediate density. In SPS, the network learns the integrated path-space balance rather than explicitly enforcing a pointwise duality relation between the two drifts. Thus, the learned drifts need not exactly satisfy Nelson's duality relation. A detailed derivation of the Gaussian kernel ratio and its continuous-time limit is provided in Appendix~\ref{app:duality_relation}.

\subsection{Independence Metropolis--Hastings for SPS}
Due to the discrepancy between the learned model distribution and the true physical distribution at the optimal convergence point, we employ the Independence  Metropolis--Hastings (IMH) algorithm to ensure exact sampling. We construct a Markov chain using the generated samples and apply an IMH correction. By imposing detailed balance on the chain, we guarantee that the final samples strictly adhere to the target physical distribution, provided that the SPS proposal covers the full support of the target distribution; evidence that this is the case in practice, in particular that both $Z_2$-related sectors are populated in the broken phase, is presented in Appendix~\ref{app:C}.

IMH operates by proposing a new configuration $\mathbf{s}'$ from the current state $\mathbf{s}$ via a proposal distribution $T(\mathbf{s}\to \mathbf{s}')$. Let $\tau(\mathbf{s}_{0}\to \mathbf{s}_{1} \to \cdots \to \mathbf{s}_{T})$ and $\tau'(\mathbf{s}_{0}^{'}\to \mathbf{s}_{1}^{'} \to \cdots \to \mathbf{s}_{T}^{'})$ denote two independent trajectories of states. The update is then accepted or rejected with probability
\begin{equation}
P_{\text{accept}}(\mathbf{s}_{T}' | \mathbf{s}_{T}) = \min\left(1, \frac{T(\mathbf{s}_{T}' \to \mathbf{s}_{T}) \tilde{\pi}^\ast(\mathbf{s}_{T}')}{T(\mathbf{s}_{T} \to \mathbf{s}_{T}') \tilde{\pi}^\ast(\mathbf{s}_{T})}\right).
\end{equation}
Detailed balance is extended to trajectory-based balance via multi-step transitions, ensuring equilibrium over entire trajectories rather than single steps.

\begin{figure}[htbp]
\centering
\subfloat[Reverse transition $T(\mathbf{s}_{T}' \to \mathbf{s}_{T})$]{
\label{fig:T_forward}
\includegraphics[height=6.5cm,width=4.9cm]{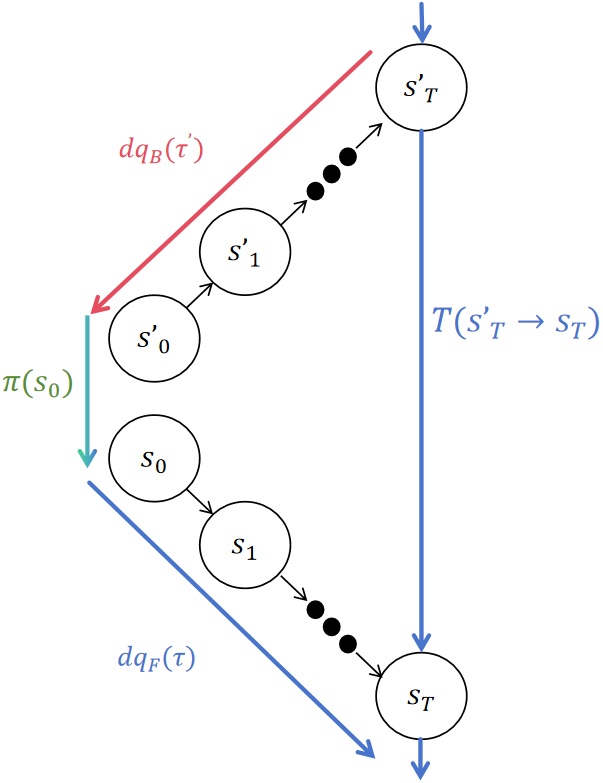}}
\subfloat[Forward transition $T(\mathbf{s}_{T} \to \mathbf{s}_{T}')$]{
\label{fig:T_reverse}
\includegraphics[height=6.5cm,width=4.9cm]{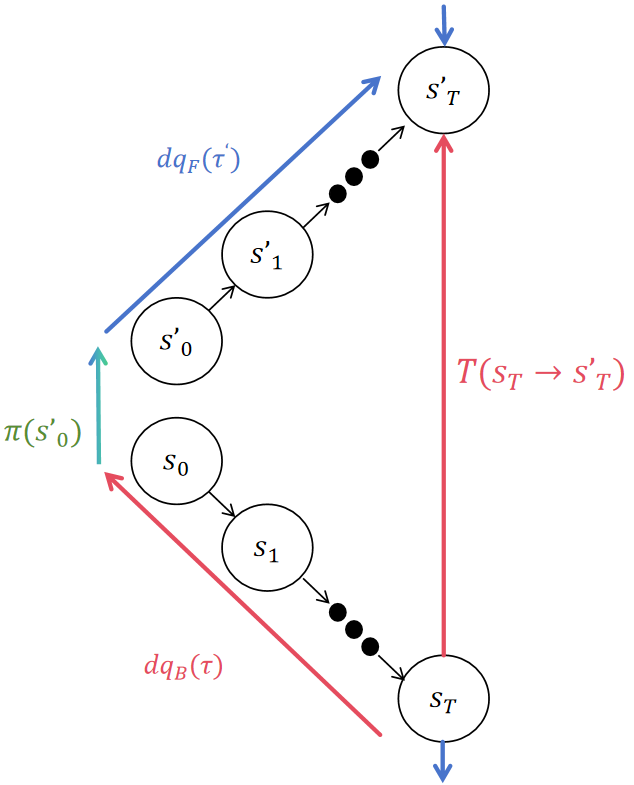}}
\subfloat[Ratio $\frac{T(\mathbf{s}_{T}' \to \mathbf{s}_{T})}{T(\mathbf{s}_{T} \to \mathbf{s}_{T}')}$]{
\label{fig:T_total}
\includegraphics[height=6.5cm,width=4.9cm]{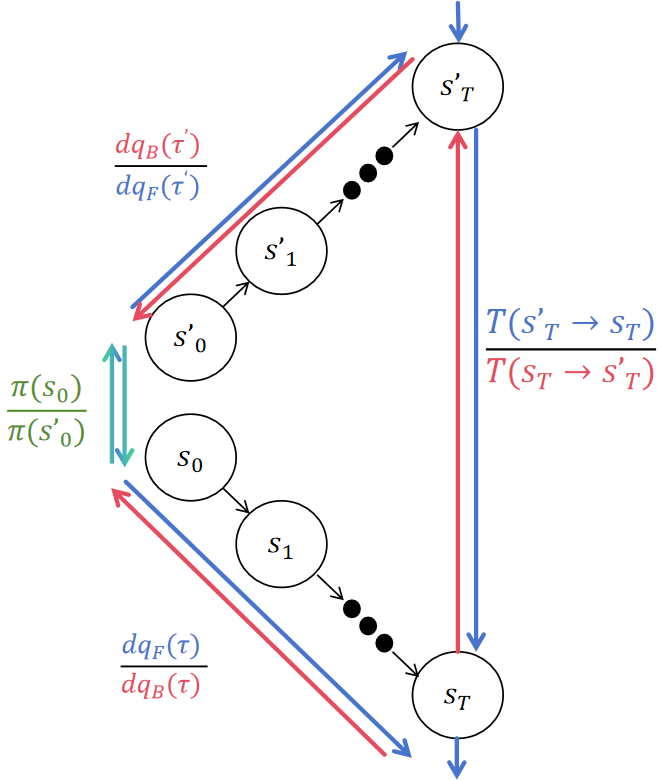}}
\caption{
Visualization of the trajectory balance principle within the IMH algorithm. \textbf{(a)} depicts the reverse probability $T(\mathbf{s}_{T}' \to \mathbf{s}_{T})$, representing the generative process from the proposed state $\mathbf{s}_{T}'$ to the current state $\mathbf{s}_{T}$. This trajectory is constructed through the sequential application of the reverse process. \textbf{(b)} illustrates the forward transition $T(\mathbf{s}_{T} \to \mathbf{s}_{T}')$, showing the forward generative path from $\mathbf{s}_{T}$ to $\mathbf{s}_{T}'$. \textbf{(c)} synthesizes the forward and reverse processes via the ratio of the probabilities.
}
\label{fig:Metropolis}
\end{figure}

Fig.~\ref{fig:T_forward} illustrates the reverse transition $T(\mathbf{s}_{T}' \to \mathbf{s}_{T})$. In the SPS framework, the reverse transition characterizes the generative trajectory starting from the proposed state $\mathbf{s}_{T}'$ and evolving to the current state $\mathbf{s}_{T}$. This is defined as the product of reverse transition probabilities along the path.  Conversely, Fig.~\ref{fig:T_reverse} depicts the forward probability $T(\mathbf{s}_{T} \to \mathbf{s}_{T}')$, which represents the forward generative process from the current state $\mathbf{s}_{T}$ to the proposed state $\mathbf{s}_{T}'$. This probability is derived from the product of forward transition probabilities along the corresponding path. Fig.~\ref{fig:T_total} integrates the previous two components by representing the ratio $T(\mathbf{s}_{T}' \to \mathbf{s}_{T}) / T(\mathbf{s}_{T} \to \mathbf{s}_{T}')$. This ratio is pivotal in the acceptance probability calculation of the enhanced IMH algorithm, as formalized in Eq.~\eqref{eq:IMH_acceptance}. The ratio ensures that the Markov chain equilibrium property, when extended to entire trajectories, is maintained by properly weighting the forward and reverse transition between states. 

The acceptance probability for the proposed state \( \mathbf{s}_{T}^{'} \) starting from $\mathbf{s}_{T}$ incorporates the full trajectory balance,
\begin{equation}
	\begin{aligned}
		P_{\text{accept}}(\mathbf{s}_{T}^{'} | \mathbf{s}_{T}) &= \min\left(1, \frac{\tilde{\pi}^\ast(\mathbf{s}_{T}^{'}) T(\mathbf{s}_{T}^{'} \to \mathbf{s}_{T})}{\tilde{\pi}^\ast(\mathbf{s}_{T}) T(\mathbf{s}_{T} \to \mathbf{s}_{T}^{'})} \right)\\
		&=\min\left(1, \frac{\tilde{\pi}^\ast(\mathbf{s}_{T}^{'})}{\tilde{\pi}^\ast(\mathbf{s}_{T})}\frac{\prod_{i=0}^{T-1} q_{\mathrm{B}}(\mathbf{s}_{i}^{'}|\mathbf{s}_{i+1}^{'})}{ \prod_{i=0}^{T-1} q_{\mathrm{F}}(\mathbf{s}_{i+1}^{'}|\mathbf{s}_{i}^{'})\,\pi_0(\mathbf{s}^{'}_0)}\frac{\prod_{i=0}^{T-1} q_{\mathrm{F}}(\mathbf{s}_{i+1}|\mathbf{s}_{i})\,\pi_0(\mathbf{s}_0)}{ \prod_{i=0}^{T-1} q_{\mathrm{B}}(\mathbf{s}_{i}|\mathbf{s}_{i+1})}\right).
	\end{aligned}
    \label{eq:IMH_acceptance}
\end{equation}

\section{SPS for Lattice \texorpdfstring{$\phi^4$}{phi\string^4} Scalar Fields}
\label{sec:3}

In Euclidean lattice field theory, the two-dimensional $\phi^4$ theory is defined by the scalar field $\phi_x$ residing on the sites $x$ of a square lattice $\Lambda$. The lattice action (or Hamiltonian in the statistical mechanics context) is given by
\begin{equation}
S(\phi) 
=\sum_{x\in\Lambda}
\bigg(-2\kappa\sum_{\mu=1}^{d} \phi_x\phi_{x+\hat{\mu}}
+ (1-2\lambda)\phi_{x}^2
+ \lambda\phi_{x}^4
\bigg),
\label{eq:action}
\end{equation}
where $\kappa$ is the hopping parameter and $\lambda \geq 0$ is the quartic (bare) coupling.  For a fixed coupling, the infinite‑volume limit exhibits a continuous phase transition as $\kappa$ increases.   When $\kappa$ surpasses the critical value $\kappa_c$ as $L\to \infty$, the system changes from a symmetric (disordered) phase, characterized by a single‑peak distribution of the field, to a broken‑symmetry (ordered) phase, characterized by a double‑peak distribution, signaling spontaneous symmetry breaking.

We focus on three standard quantities with volume $|\Lambda|=N_x\times N_\tau = L\times 8$, which follows the finite-temperature lattice geometry used in Ref.~\cite{Nicoli:2020njz}. In this setup, the second direction is interpreted as the Euclidean temporal direction, whose finite extent $N_\tau=8$ fixes the temperature scale through $T=1/(N_\tau a)$, while the spatial extent $N_x=L$ is varied. This allows us to study the dependence of the observables and sampling performance on the spatial volume at fixed temporal extent, and hence at fixed temperature in lattice units.

\begin{itemize}
    \item \textbf{Magnetization $M$} and \textbf{Absolute Magnetization $|M|$}:
    \begin{equation}
        m = \frac{1}{|\Lambda|}\sum_{x\in\Lambda} \phi(x),\quad
        M = \langle m\rangle, \quad |M| = \langle |m|\rangle.
    \end{equation}
    The absolute magnetization serves as an order parameter for detecting spontaneous symmetry breaking in finite volume, since the ordinary magnetization $\langle m\rangle$ may vanish due to the underlying $\mathbb{Z}_2$ symmetry.
    \item \textbf{Susceptibility $\chi$}:
    \begin{equation}
        \chi =|\Lambda|\left(
        \langle m^2\rangle - \langle |m|\rangle^2
        \right).
    \end{equation}
    the magnetic susceptibility describes the fluctuations of the field. It is large near the critical point, where the correlation length diverges.
    \item \textbf{Free energy density} $F$:
    \begin{equation}
        F = -\frac{1}{|\Lambda|}\log Z =  - \frac{1}{|\Lambda|}\log \int D\phi\, \tilde{\pi}^\ast(\phi)  = - \frac{1}{|\Lambda|}\log \mathbb{E}_{dq_{\mathrm{F}}}\!\left[ w(\phi;\tau)\right],
        \label{eq:free_energy_importance}
    \end{equation}
    with corresponding reweighting factor,
    \begin{equation}
    w(\phi;\tau) = \frac{\tilde{\pi}^\ast(\phi)}{q_\theta(\phi;\tau)} =  e^{\log\tilde{\pi}^\ast(\phi) - \log q_\theta(\phi;\tau)},
    \label{eq:reweight_ensemble}
\end{equation}
and the path distribution for $\phi$,
\begin{equation}
    q_\theta(\phi;\tau) = q_\theta(\bm{s}_T;\tau) \equiv \frac{\prod_{i=0}^{T-1} q_{\mathrm{F}}(\bm{s}_{i+1}\mid \bm{s}_i)\,\pi_0(\bm{s}_0)}{\prod_{i=0}^{T-1} q_{\mathrm{B}}(\bm{s}_{i}\mid \bm{s}_{i+1})}.
\end{equation}
which is defined using Eq.~\eqref{eq:tb_ratio_general}. Since $q_\theta(\phi;\tau)$ is in general path-dependent at finite training, the weight $w(\phi;\tau)$ is a function of the full trajectory rather than of the endpoint alone; the reweighted estimator in Eq.~\eqref{eq:free_energy_importance} nevertheless remains unbiased in expectation over the forward path measure $dq_{\mathrm{F}}$.
\end{itemize}

Before presenting the results of our method, we emphasize that all experiments are conducted in a fully dataless setting: no training samples from HMC or other samplers are used, and the model is trained solely by evaluating the target action $S(\phi)$. A separate network is trained for each pair of hopping parameter and lattice size $(\kappa, L)$ considered; no conditioning across couplings or volumes is employed. The network architecture, training and generation are described in Appendix~\ref{sec:Network}. The number of steps used in training ($T_{\rm train}$) and generation ($T_{\rm gen}$) can be different, using time-embedding. The results shown below use $T_{\rm train}=250$ and $T_{\rm gen}=2500$.

In the following results, as for comparison, HMC simulations are performed on lattices with $L=16,32,48,64$, and periodic boundary conditions. The quartic coupling is fixed at $\lambda=0.022$, while the hopping parameter is scanned over $\kappa\in[0.20,0.30]$ with spacing $\Delta\kappa=0.01$.  In the infinite-volume limit ($L\to \infty$), the critical value $\kappa_c \approx 0.239$ for this value of the coupling. For each parameter set, we first discard $50,000$ trajectories for thermalization. We then generate $2\times10^6$ HMC trajectories using leapfrog integration with step size $\Delta t=0.01$ and $100$ leapfrog steps per trajectory. Measurements are recorded every $500$ trajectories, yielding approximately $4,000$ decorrelated configurations for estimating the observables. For the finite $L\times 8$ lattices considered here, the observables show a rapid crossover and a susceptibility peak near $\kappa\approx 0.27$. We refer to this region as the pseudocritical region for the finite-volume setup. A determination of the infinite-volume critical coupling would require a dedicated finite-size scaling analysis and is outside the scope of the present benchmark.

We highlight that, at least for the simulated lattice sizes, our algorithm does not appear to suffer from mode collapse (see Appendix~\ref{app:C} for example histograms of the magnetisation in the pseudocritical region).

\subsection{Absolute Magnetization}
\begin{figure}[htbp]
\centering
\subfloat[$L=16$]{
\label{fig:magent16}
\includegraphics[width=0.49\linewidth]{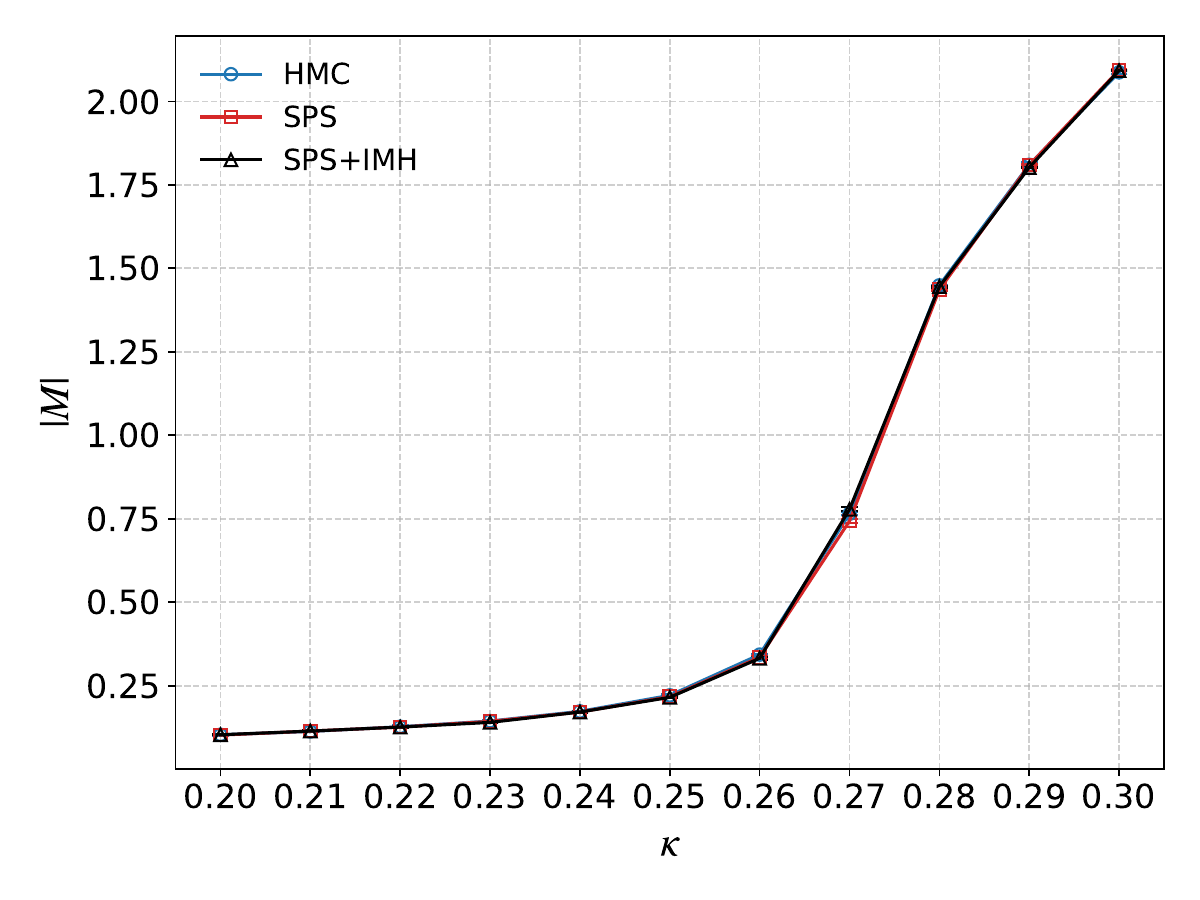}}
\subfloat[$L=32$]{
\label{fig:magent32}
\includegraphics[width=0.49\linewidth]{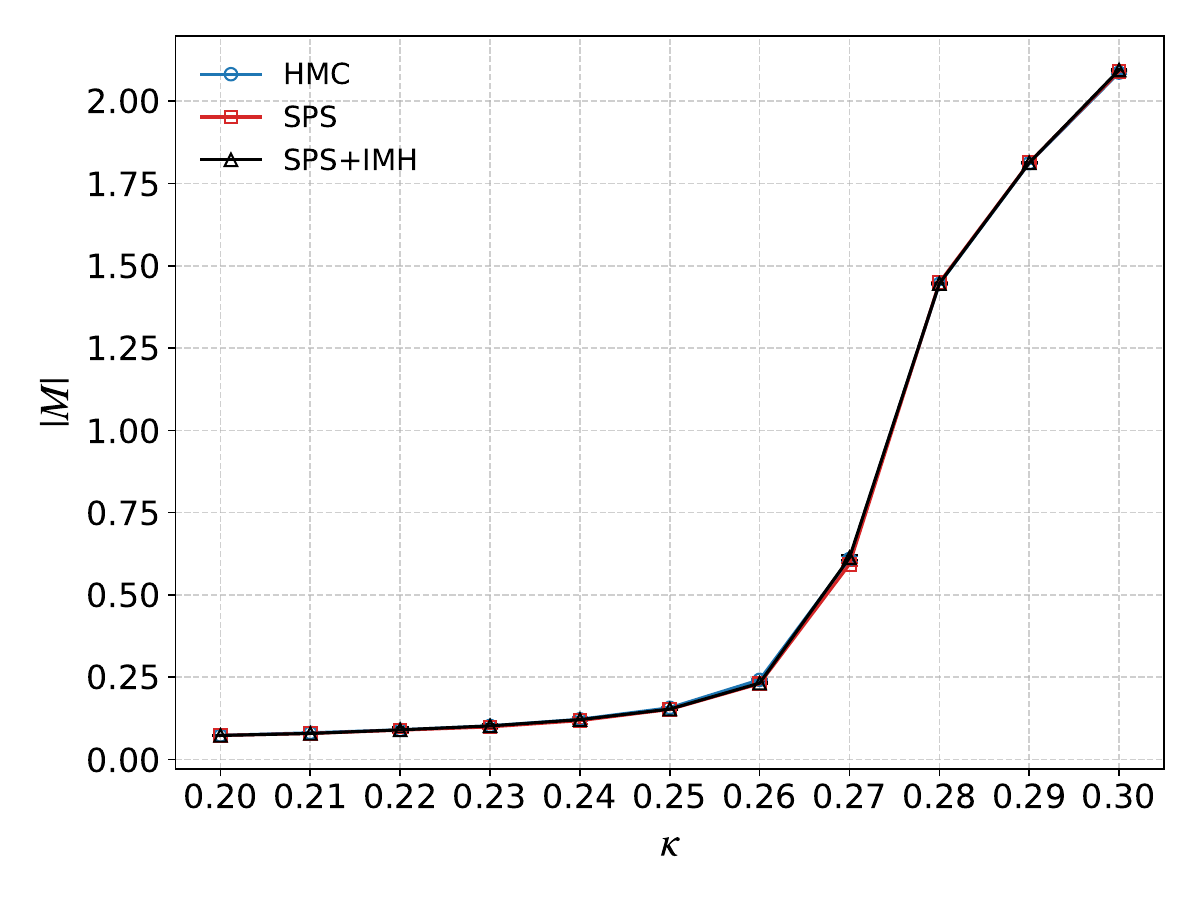}}\\
\subfloat[$L=48$]{
\label{fig:magent48}
\includegraphics[width=0.49\linewidth]{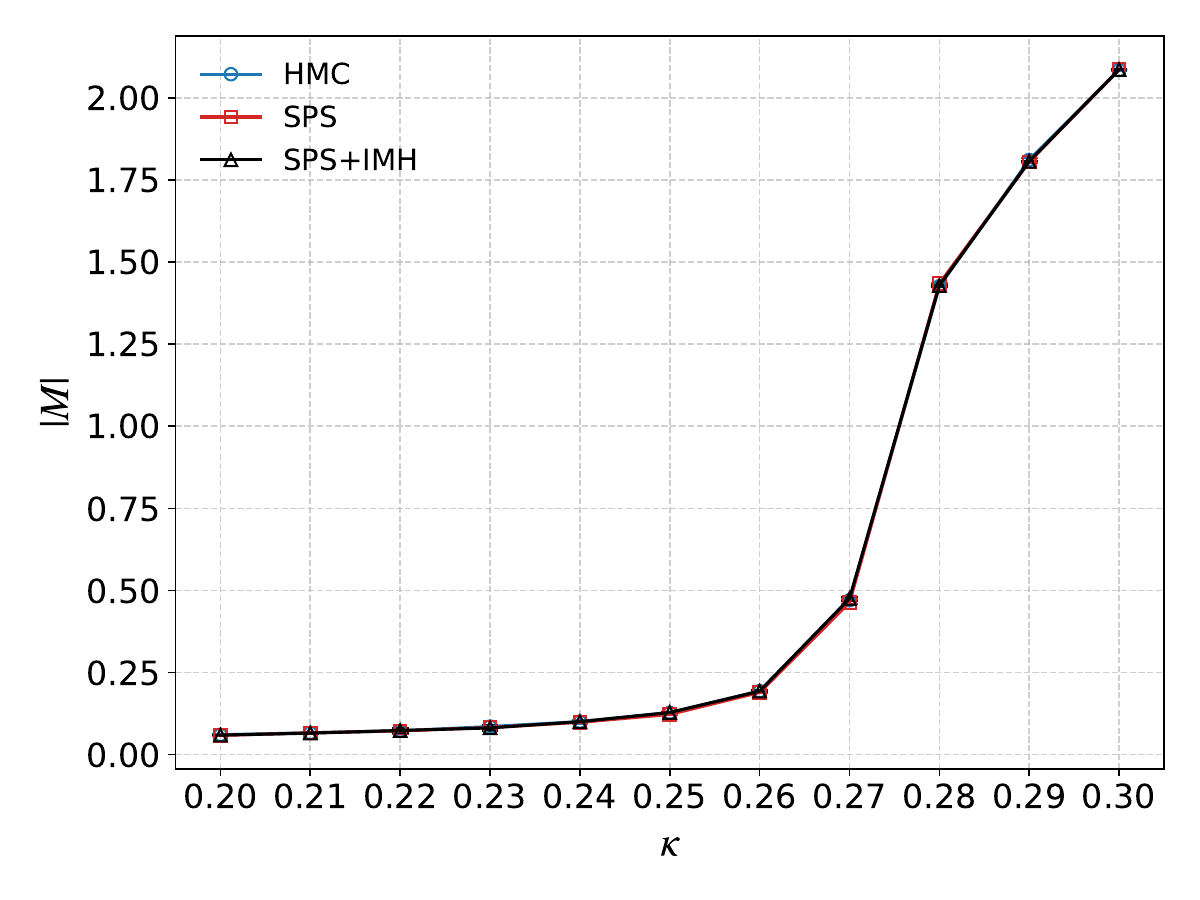}}
\subfloat[$L=64$]{
\label{fig:magent64}
\includegraphics[width=0.49\linewidth]{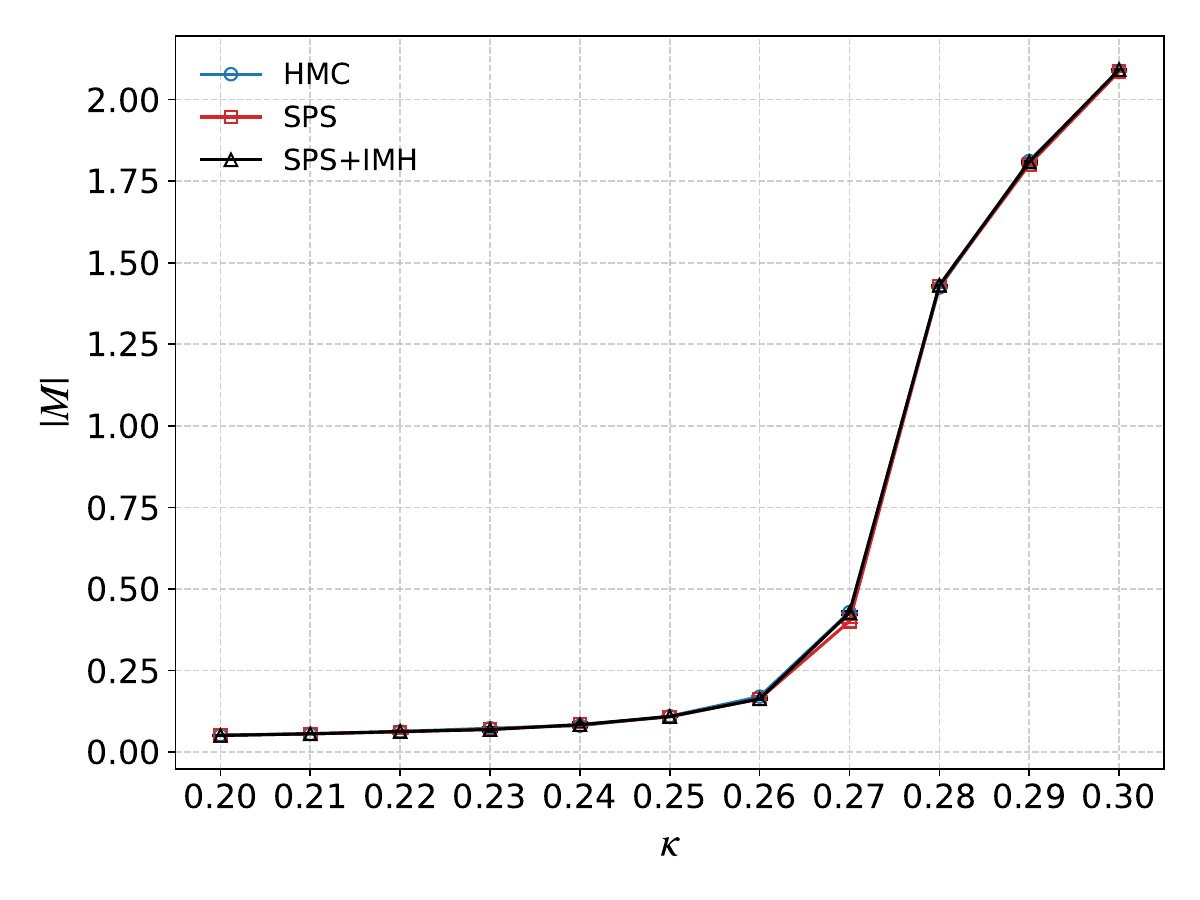}}
\caption{Absolute magnetization $|M|$ from HMC (blue), SPS (red), and SPS$+$IMH(black).
Note that the data are essentially indistinguishable, see also Table~\ref{tab:magnetization}. }
\label{fig:magnet}
\end{figure}

In Fig.~\ref{fig:magnet}, we present the absolute magnetization as a function of the hopping parameter $\kappa$ for different lattice sizes $L$. The blue data points represent the HMC benchmark results, the red data points denote the uncorrected SPS results, and the black data points correspond to SPS with the Independence Metropolis--Hastings correction. We show the uncorrected SPS results to assess the quality of the learned proposal distribution before the correction, while the SPS$+$IMH results demonstrate the effect of the IMH correction in enforcing exact sampling from the target distribution. These data points exhibit good agreement across the entire $\kappa$ range, demonstrating that the field configurations generated by SPS are in good agreement with those produced by HMC.
Furthermore, a distinct jump in magnetization is observed near $\kappa = 0.27$ for all curves, indicating that SPS successfully captures the phase transition signal.

\begin{table}[htbp]
\centering
 \scalebox{0.95}{
\setlength{\tabcolsep}{8pt} 
\begin{tabular}{c c S[table-format=2.4(4)] S[table-format=2.4(4)] S[table-format=2.4(4)] S[table-format=2.4(4)]}
\toprule
$\kappa$ & Method & {$L=16\qquad$} & {$32\qquad$} & {$48\qquad$} & {$64\qquad$} \\
\midrule

0.20 & HMC & 0.103(3) & 0.073(1) & 0.060(1) & 0.052(1) \\
     & SPS & 0.102(1) & 0.073(1) & 0.058(1) & 0.051(1) \\
     & SPS$+$IMH & 0.103(2) & 0.073(1) & 0.059(1) & 0.051(1) \\
\hline

0.21 & HMC & 0.113(1) & 0.081(1) & 0.066(1) & 0.056(1) \\
     & SPS & 0.114(2) & 0.079(1) & 0.066(1) & 0.056(1) \\
     & SPS$+$IMH & 0.114(2) & 0.079(2) & 0.065(1) & 0.056(1) \\
\hline

0.22 & HMC & 0.127(5) & 0.090(4) & 0.072(3) & 0.063(1) \\
     & SPS & 0.126(2) & 0.089(7) & 0.072(8) & 0.062(7) \\
     & SPS$+$IMH & 0.126(2) & 0.090(7) & 0.073(8) & 0.063(2) \\
\hline

0.23 & HMC & 0.144(2) & 0.102(5) & 0.085(5) & 0.073(4) \\
     & SPS & 0.143(2) & 0.099(7) & 0.082(8) & 0.070(7) \\
     & SPS$+$IMH & 0.140(2) & 0.102(7) & 0.081(8) & 0.069(7) \\
\hline

0.24 & HMC & 0.173(1) & 0.122(6) & 0.100(6) & 0.081(5) \\
     & SPS & 0.171(2) & 0.118(9) & 0.098(2) & 0.084(9) \\
     & SPS$+$IMH & 0.171(3) & 0.121(2) & 0.100(2) & 0.084(8) \\
\hline

0.25 & HMC & 0.221(3) & 0.157(1) & 0.127(1) & 0.110(1) \\
     & SPS & 0.217(2) & 0.152(2) & 0.122(2) & 0.108(1) \\
     & SPS$+$IMH & 0.215(3) & 0.153(3) & 0.128(2) & 0.109(1) \\
\hline

0.26 & HMC & 0.343(3) & 0.242(2) & 0.193(3) & 0.167(2) \\
     & SPS & 0.335(4) & 0.230(2) & 0.189(3) & 0.163(2) \\
     & SPS$+$IMH & 0.332(5) & 0.232(4) & 0.193(3) & 0.163(3) \\
\hline

0.27 & HMC & 0.765(5) & 0.609(9) & 0.470(3) & 0.429(3) \\
     & SPS & 0.743(6) & 0.593(5) & 0.463(4) & 0.401(4) \\
     & SPS$+$IMH & 0.778(6) & 0.613(7) & 0.475(6) & 0.427(6) \\
\hline

0.28 & HMC & 1.449(2) & 1.445(2) & 1.427(2) & 1.425(2) \\
     & SPS & 1.436(3) & 1.449(2) & 1.435(2) & 1.429(2) \\
     & SPS$+$IMH & 1.445(3) & 1.446(2) & 1.429(3) & 1.431(2) \\
\hline

0.29 & HMC & 1.810(7) & 1.812(2) & 1.812(7) & 1.812(6) \\
     & SPS & 1.809(2) & 1.814(2) & 1.806(2) & 1.801(2) \\
     & SPS$+$IMH & 1.802(2) & 1.813(2) & 1.8075(2) & 1.809(2)  \\
\hline

0.30 & HMC & 2.087(6) & 2.087(5) & 2.086(2) & 2.088(2) \\
     & SPS & 2.094(2) & 2.090(2) & 2.087(1) & 2.086(1) \\
    & SPS$+$IMH & 2.092(2) & 2.095(2) & 2.087(1) & 2.092(1) \\

\bottomrule
\end{tabular}
}
\caption{
Absolute magnetization ($|M|$) as a function of $\kappa$ for lattice sizes $L=16,32,48,64$.
Results from HMC, SPS, and SPS$+$IMH are shown in separate rows.}
\label{tab:magnetization}
\end{table}

Table~\ref{tab:magnetization} presents the numerical values of absolute  magnetization obtained from both SPS and HMC simulations. The majority of SPS results agree with the HMC benchmarks within the $1\sigma$ statistical error range. At specific parameter points near and beyond the pseudocritical region, namely $\kappa =0.26 - 0.28$, the agreement extends to within $1.5\sigma$ for certain lattice sizes.

The deviation can be attributed to the system approaching and eventually entering the symmetry-broken phase, where long-range correlations and enhanced fluctuations make the field distribution more difficult to approximate accurately. In the vicinity of the pseudocritical point, the absolute magnetization distribution exhibits mild non-Gaussian features due to enhanced critical fluctuations. By contrast, at $\kappa = 0.27$, where the system has entered the broken-symmetry phase, the magnetization distribution develops a bimodal structure, with a nonnegligible probability density remaining in the region between the two separated modes. This nontrivial topology poses a challenge for KL-divergence-based SPS training (see Appendix~\ref{app:C}).

The application of the IMH correction significantly reduces these discrepancies, bringing the SPS$+$IMH results into agreement with HMC across all parameter values. This observation underscores the critical importance of the IMH correction step in ensuring exact sampling, particularly in regions where the target distribution exhibits nontrivial topological features.

\subsection{Susceptibility}
\begin{figure}[htbp]
\centering
\subfloat[$L=16$]{
\label{fig:susc16}
\includegraphics[width=0.49\linewidth]{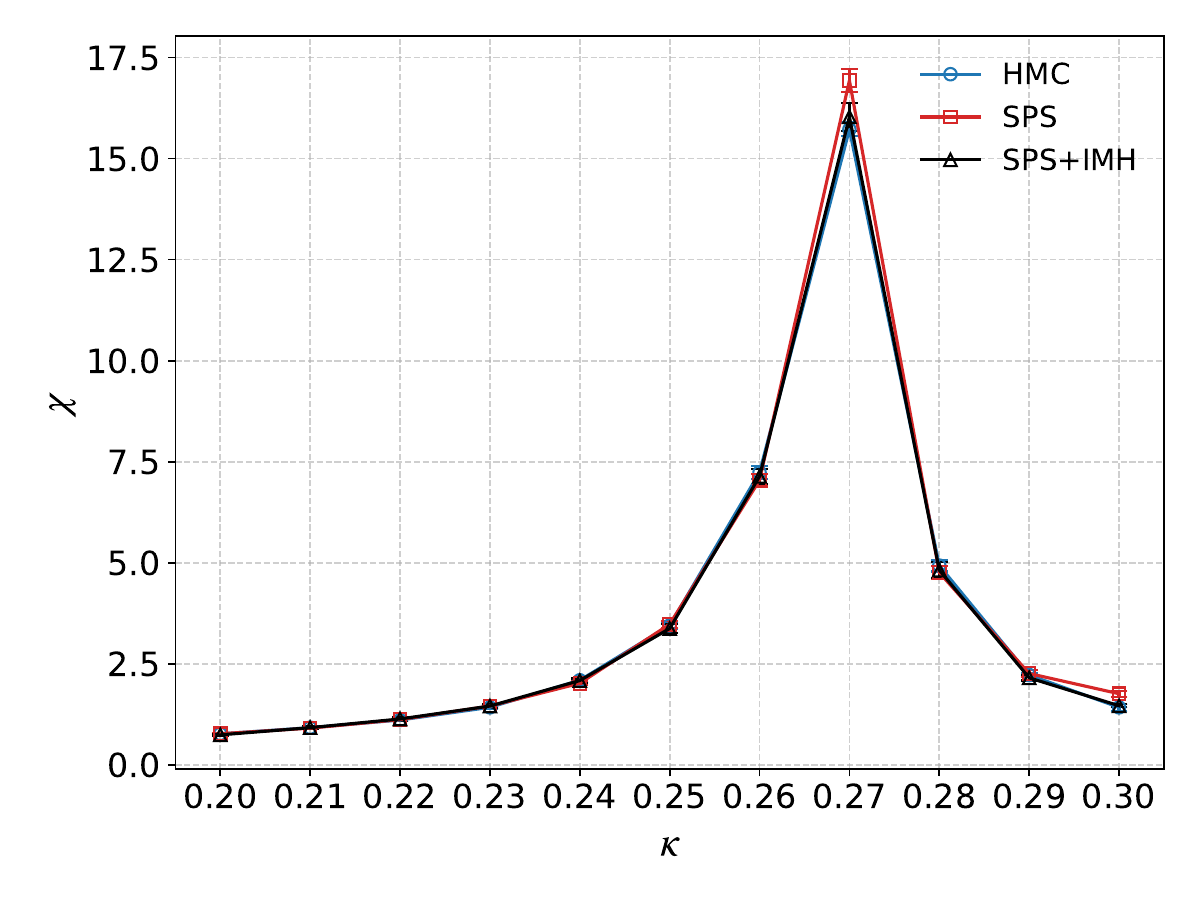}}
\subfloat[$L=32$]{
\label{fig:susc32}
\includegraphics[width=0.49\linewidth]{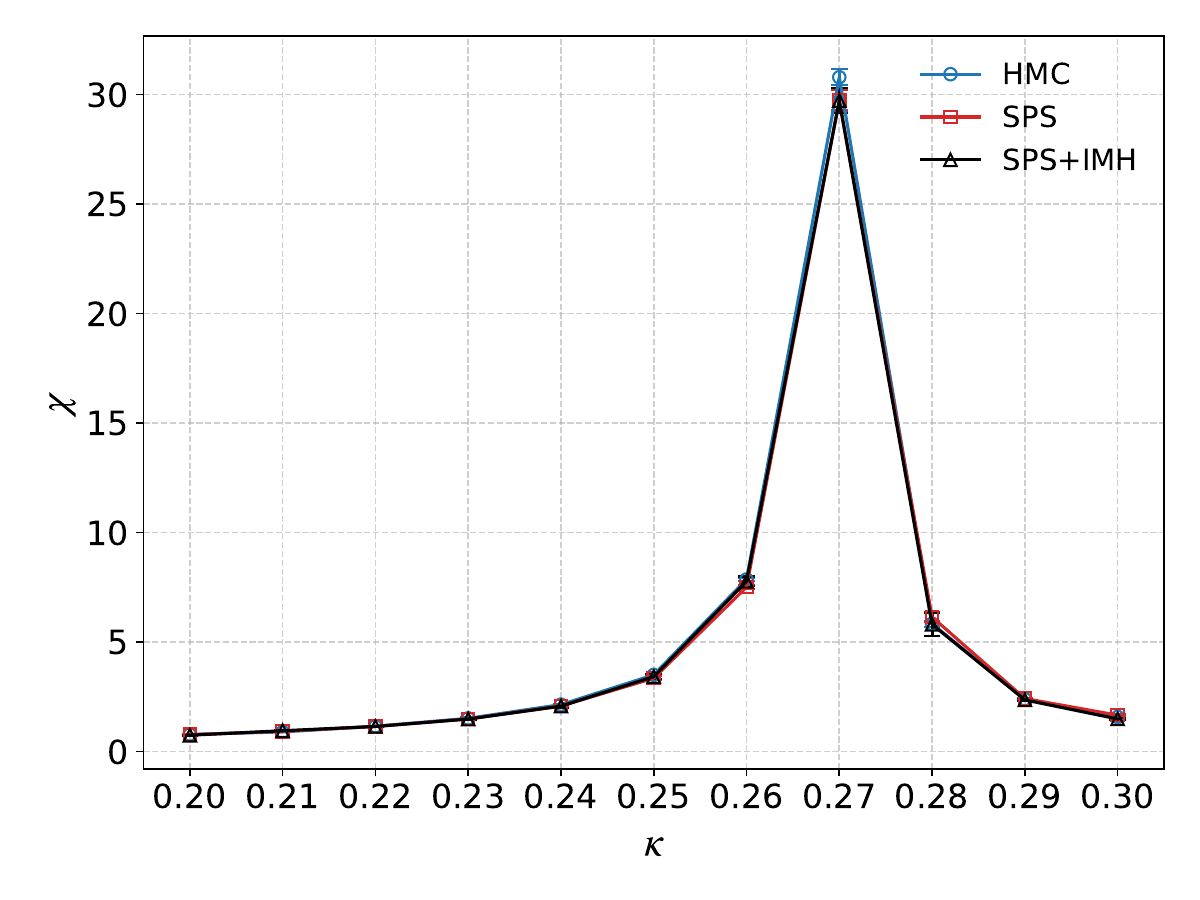}}\\
\subfloat[$L=48$]{
\label{fig:susc48}
\includegraphics[width=0.49\linewidth]{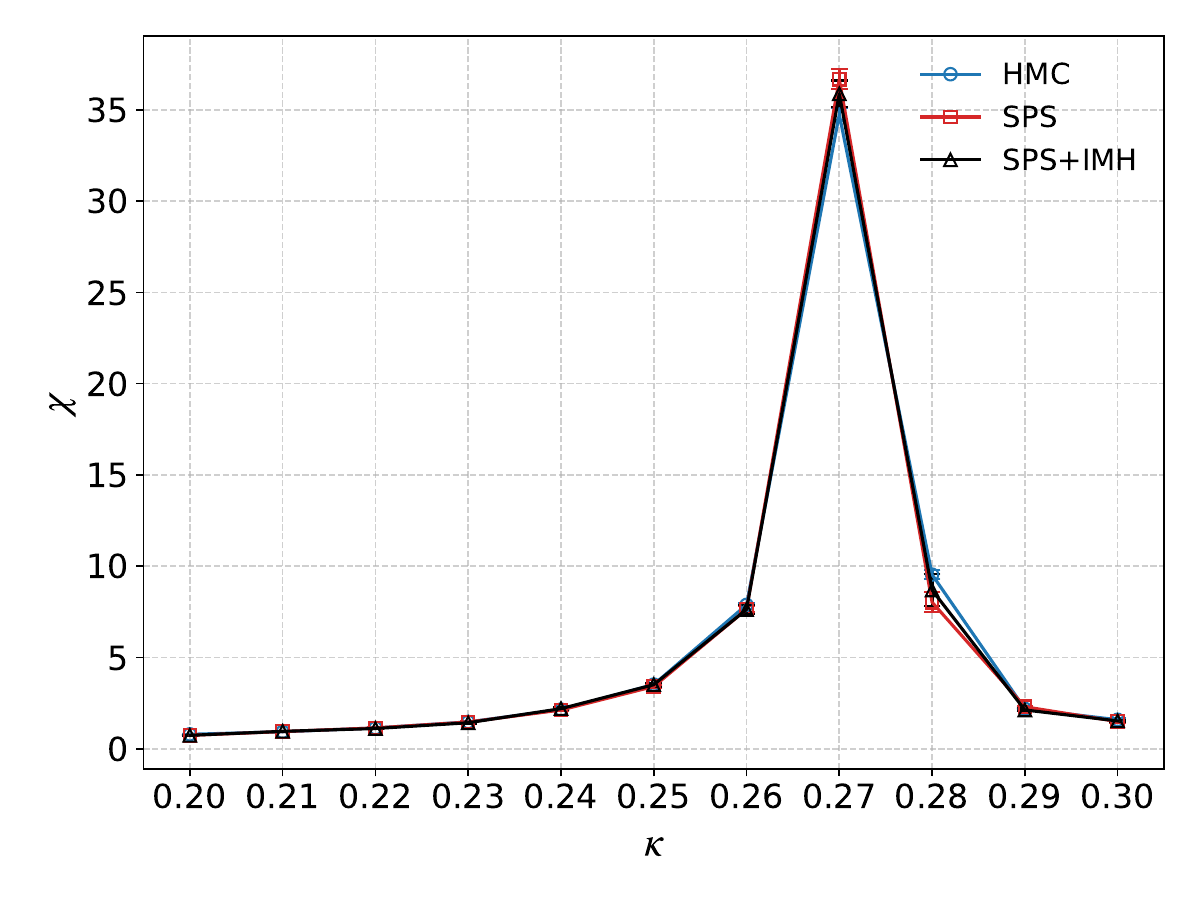}}
\subfloat[$L=64$]{
\label{fig:susc64}
\includegraphics[width=0.49\linewidth]{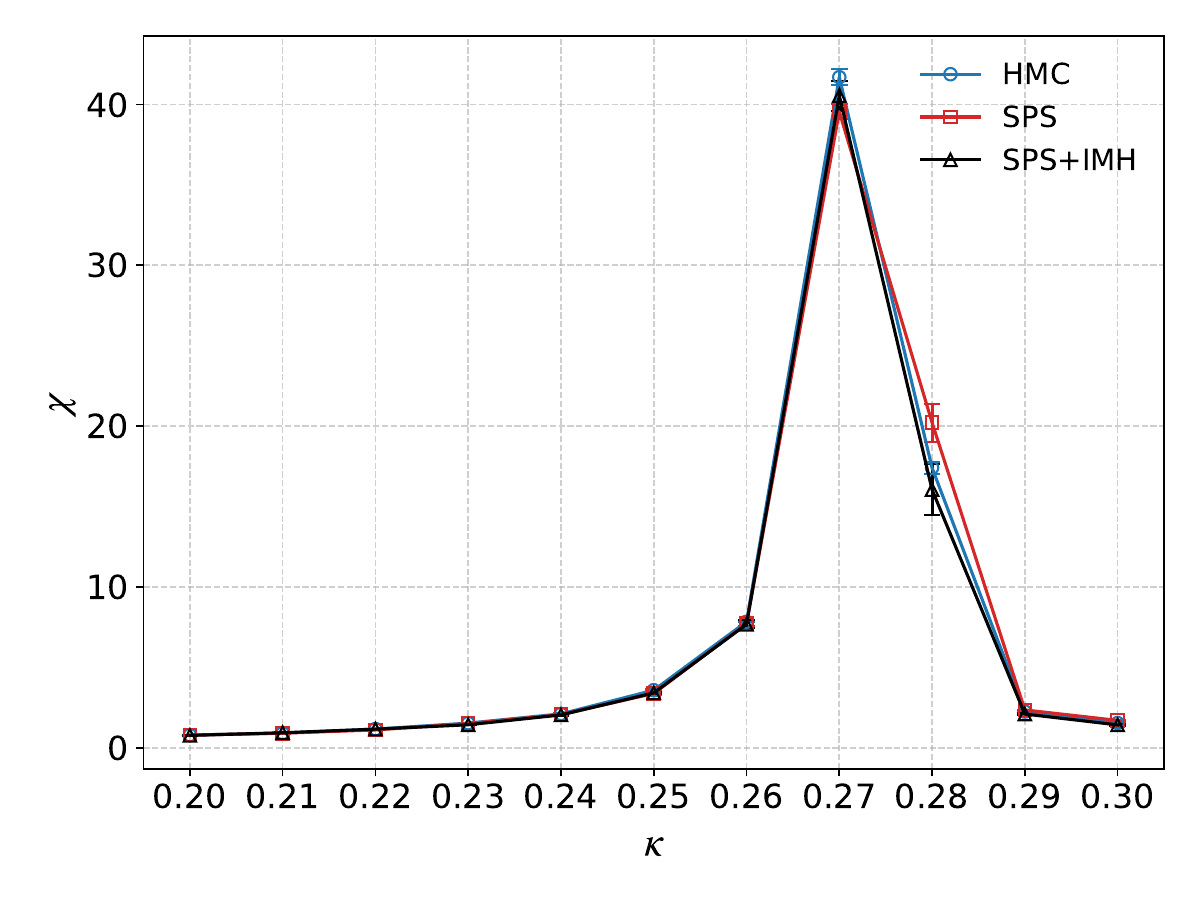}}
\caption{Susceptibility $\chi$ from HMC (blue), SPS (red), and SPS$+$IMH (black)}
\label{fig:susc}
\end{figure}

Fig.~\ref{fig:susc} presents the susceptibility $\chi$ as a function of the hopping parameter $\kappa$ for various lattice sizes $L$.  The corresponding numerical data are provided in Table~\ref{tab:susceptibility_data}.  Deep in the symmetric phase ($\kappa < 0.24$), both HMC and SPS yield consistently small susceptibility values, consistent within the $1\sigma$ statistical error margin, and indicating minimal field fluctuations.  As the system approaches the pseudocritical point, the susceptibility rises sharply, culminating in a pronounced peak at $\kappa \sim 0.27$, which signifies the onset of long-range correlations characteristic of a continuous phase transition. For $\kappa \gtrsim 0.27$, a notable deviation emerges between the uncorrected SPS results and the HMC benchmarks, exceeding the $3\sigma$ confidence interval.   This discrepancy is attributed to the nonzero diffusion coefficient at the terminal time step $t=T$, which continues to drive the SPS to explore rare, high-action field configurations, affecting the tails of the distribution that are crucial for accurately capturing susceptibility. Crucially, this deviation is rectified by the IMH correction. As demonstrated by the SPS$+$IMH data in both Fig.~\ref{fig:susc} and Table~\ref{tab:susceptibility_data}, the corrected results exhibit remarkable agreement with the HMC benchmarks across all $\kappa$ values and lattice sizes, validating the essential role of the IMH correction step in ensuring exact sampling.


\begin{table}[htbp]
\centering
\scalebox{0.95}{
\setlength{\tabcolsep}{8pt} 

\begin{tabular}{c c c c c c}
\toprule
$\kappa$ & Method & $L=16$ & 32 & 48 & 64 \\
\midrule
0.20 & HMC & 0.776(5) & 0.787(7) & 0.793(10) & 0.792(10) \\
     & SPS & 0.778(21) & 0.773(13) & 0.736(17) & 0.792(19) \\
     & SPS$+$IMH & 0.750(22) & 0.750(25) & 0.736(23) & 0.772(22) \\
    \hline
0.21 & HMC & 0.928(7) & 0.900(23) & 0.933(11) & 0.917(10) \\
     & SPS & 0.916(15) & 0.931(16) & 0.944(23) & 0.908(21) \\
     & SPS$+$IMH & 0.928(30) & 0.954(32) & 0.955(23) & 0.936(31) \\
     \hline
0.22 & HMC & 1.117(28) & 1.160(11) & 1.140(14) & 1.167(16) \\
     & SPS & 1.128(30) & 1.148(20) & 1.142(26) & 1.113(26) \\
     & SPS$+$IMH & 1.143(38) & 1.150(38) & 1.110(36) & 1.162(25) \\
     \hline
0.23 & HMC & 1.431(40) & 1.520(41) & 1.472(20) & 1.536(21) \\
     & SPS & 1.462(37) & 1.491(25) & 1.462(35) & 1.506(35) \\
     & SPS$+$IMH & 1.463(48) & 1.488(48) & 1.425(50) & 1.433(30) \\
     \hline
0.24 & HMC & 2.100(51) & 2.148(23) & 2.146(27) & 2.109(26) \\
     & SPS & 2.025(48) & 2.082(33) & 2.123(46) & 2.065(49) \\
     & SPS$+$IMH & 2.093(50) & 2.077(66) & 2.197(78) & 2.045(49) \\
     \hline
0.25 & HMC & 3.455(85) & 3.505(34) & 3.517(55) & 3.582(54) \\
     & SPS & 3.486(89) & 3.354(86) & 3.410(79) & 3.380(78) \\
     & SPS$+$IMH& 3.38(10) & 3.42(11) & 3.51(11) & 3.412(72) \\
     \hline
0.26 & HMC & 7.24(15) & 7.864(70) & 7.883(80) & 7.840(40) \\
     & SPS & 7.05(16) & 7.51(11) & 7.64(17) & 7.71(17) \\
     & SPS$+$IMH & 7.14(20) & 7.78(22) & 7.63(23) & 7.68(25) \\
     \hline
0.27 & HMC & 15.72(17) & 30.79(36) & 34.74(33) & 41.65(49) \\
     & SPS & 17.93(28) & 29.74(46) & 36.67(54) & 39.55(48) \\
     & SPS$+$IMH & 16.03(35) & 29.73(59) & 35.87(73) & 40.55(93) \\
     \hline
0.28 & HMC & 4.94(15) & 5.84(14) & 9.54(24) & 17.41(39) \\
     & SPS & 4.76(16) & 6.15(24) & 8.03(55) & 20.2(12) \\
     & SPS$+$IMH & 4.82(20) & 5.81(52) & 8.70(90)  & 16.1(16) \\
     \hline
0.29 & HMC & 2.229(53) & 2.401(18) & 2.190(24) & 2.255(52) \\
     & SPS & 2.267(84) & 2.425(21) & 2.302(25) & 2.355(71) \\
     & SPS$+$IMH & 2.160(66) & 2.362(27) & 2.126(80) & 2.191(81) \\
     \hline
0.30 & HMC & 1.430(90) & 1.582(97) & 1.590(89) & 1.540(34) \\
     & SPS & 1.767(70) & 1.664(46) & 1.498(66) & 1.705(14) \\
     & SPS$+$IMH & 1.470(40) & 1.493(50) & 1.525(61) & 1.425(52) \\
\bottomrule
\end{tabular}
}
\caption{Susceptibility as a function of $\kappa$ for lattice sizes $L = 16, 32, 48, 64$. Results from HMC, SPS, and SPS$+$IMH are shown.}
\label{tab:susceptibility_data}
\end{table}

\subsection{Free Energy Density}

\begin{figure}[htbp]
\centering
\subfloat[$L=16$]{
\label{fig:free16}
\includegraphics[width=0.49\linewidth]{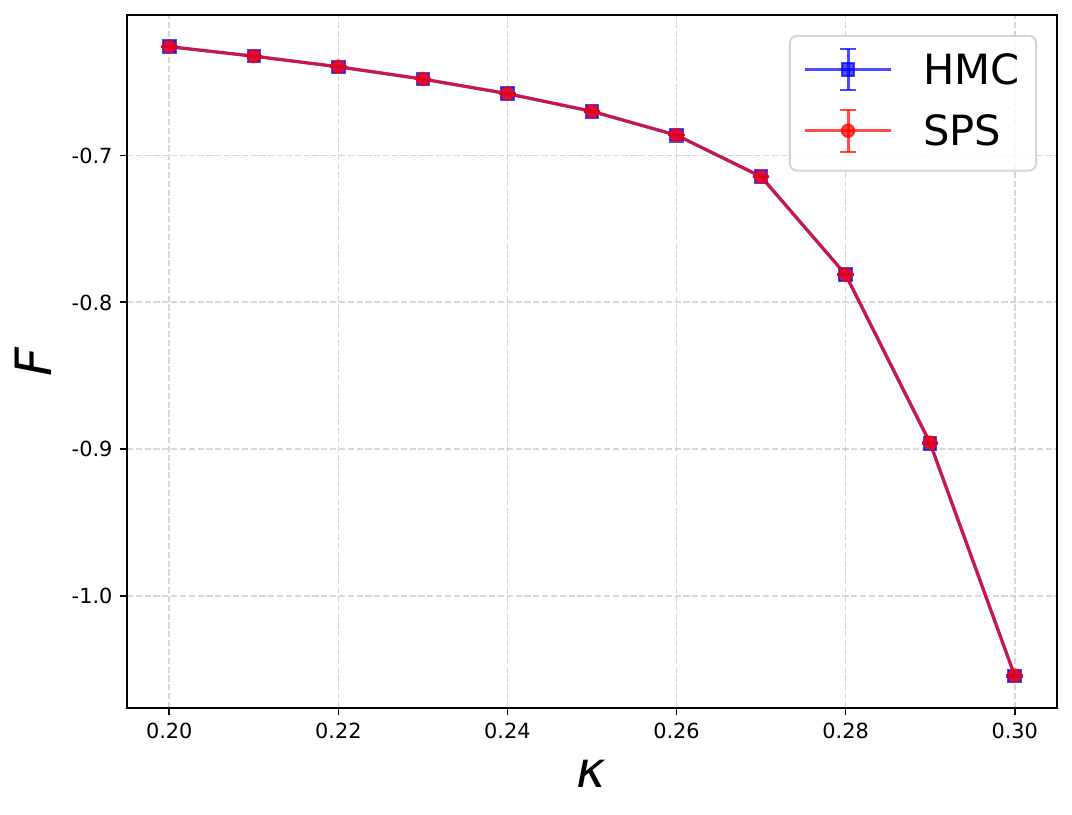}}
\subfloat[$L=32$]{
\label{fig:free32}
\includegraphics[width=0.49\linewidth]{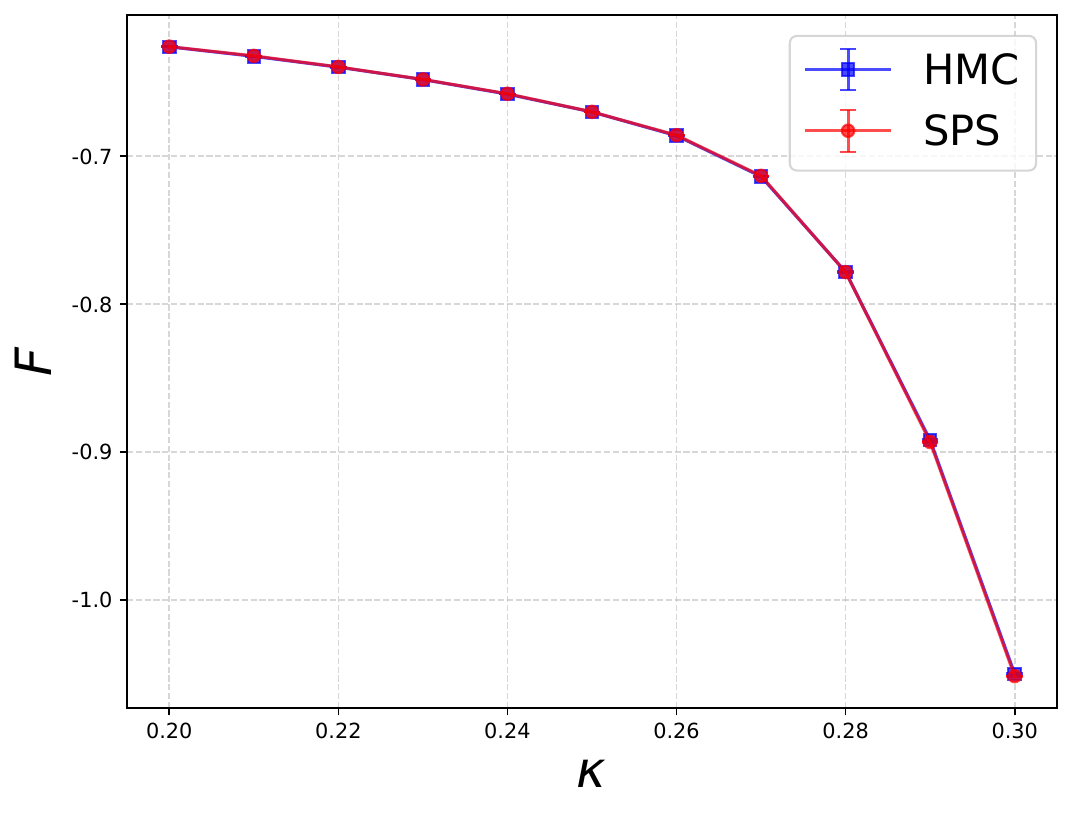}}\\
\subfloat[$L=48$]{
\label{fig:free48}
\includegraphics[width=0.49\linewidth]{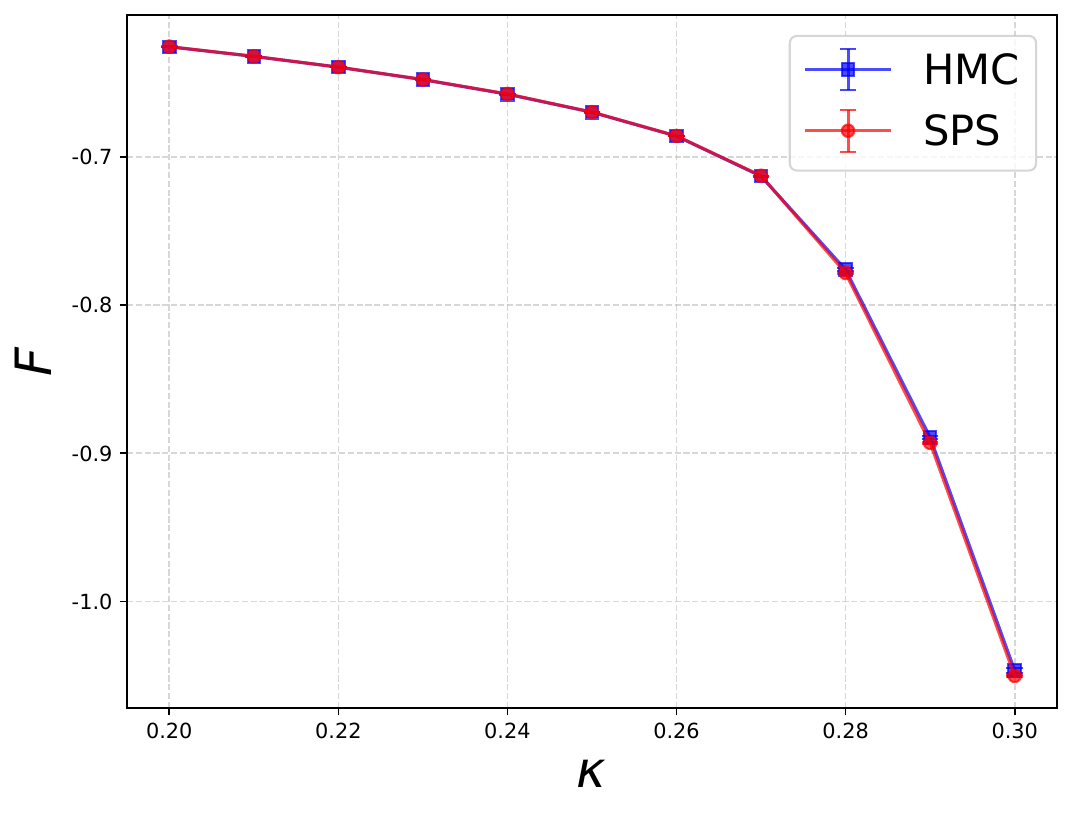}}
\subfloat[$L=64$]{
\label{fig:free64}
\includegraphics[width=0.49\linewidth]{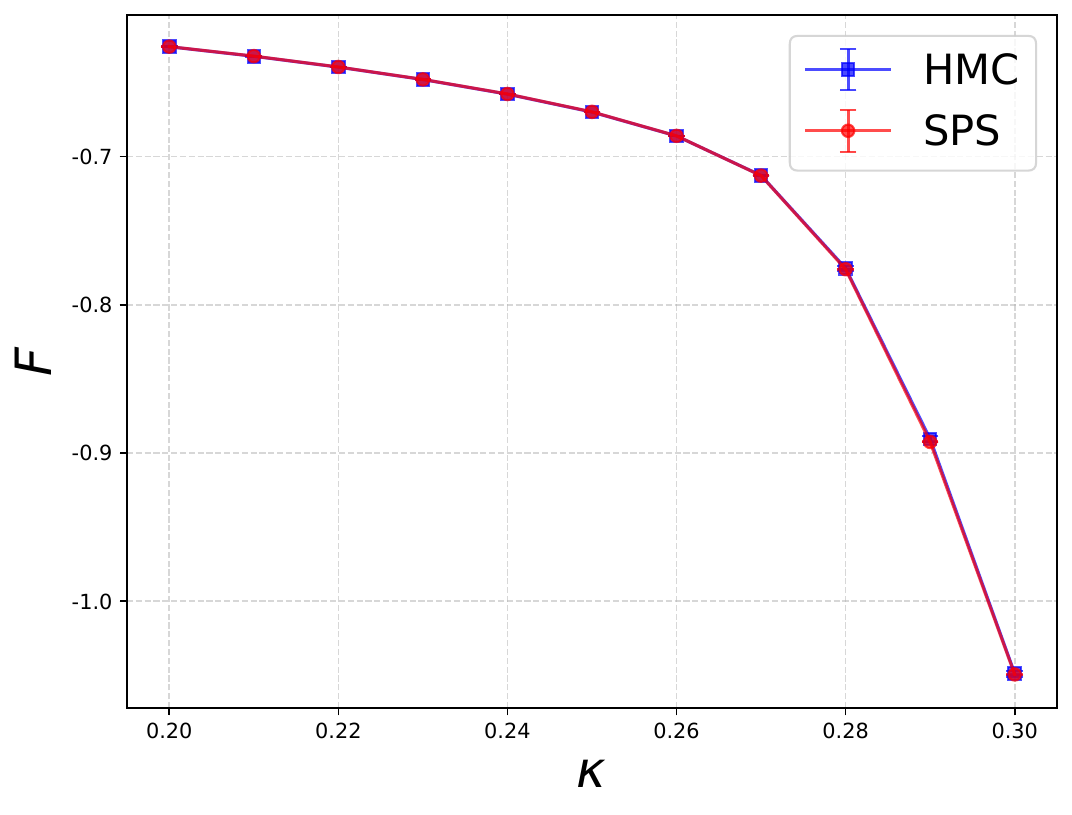}}
\caption{Free energy density $F$ from HMC (blue) and SPS (red).}
\label{fig:free}
\end{figure}

\begin{table}[htbp]
\centering
\scalebox{0.98}{
\setlength{\tabcolsep}{8pt}
\begin{tabular}{c c S[table-format=2.4(4)] S[table-format=2.4(4)] S[table-format=2.4(4)] S[table-format=2.4(4)]}
\toprule
$\kappa$ & Method & {$L=16\qquad$} & {$32\qquad$} & {$48\qquad$} & {$64\qquad$} \\
\midrule

0.20 & HMC & -0.6259(7)  & -0.6260(22) & -0.6259(16) & -0.6259(14) \\
     & SPS & -0.6259(10) & -0.6259(10) & -0.6258(8)  & -0.6258(9)  \\
\hline
0.21 & HMC & -0.6323(7)  & -0.6325(22) & -0.6323(16) & -0.6323(14) \\
     & SPS & -0.6324(11) & -0.6322(8)  & -0.6322(10) & -0.6322(7)  \\
\hline
0.22 & HMC & -0.6396(7)  & -0.6398(22) & -0.6396(16) & -0.6396(14) \\
     & SPS & -0.6395(13) & -0.6396(9)  & -0.6396(9)  & -0.6396(9)  \\
\hline
0.23 & HMC & -0.6480(7)  & -0.6482(22) & -0.6480(16) & -0.6480(14) \\
     & SPS & -0.6479(11) & -0.6481(9)  & -0.6480(9)  & -0.6479(9)  \\
\hline
0.24 & HMC & -0.6579(7)  & -0.6580(22) & -0.6577(17) & -0.6578(14) \\
     & SPS & -0.6579(12) & -0.6578(9)  & -0.6577(8)  & -0.6578(8)  \\
\hline
0.25 & HMC & -0.6700(8)  & -0.6701(22) & -0.6699(17) & -0.6700(14) \\
     & SPS & -0.6699(14) & -0.6700(12) & -0.6699(11) & -0.6698(12) \\
\hline
0.26 & HMC & -0.6862(8)  & -0.6861(22) & -0.6861(17) & -0.6861(14) \\
     & SPS & -0.6862(15) & -0.6858(11) & -0.6860(13) & -0.6860(13) \\
\hline
0.27 & HMC & -0.7142(18) & -0.7137(42) & -0.7132(35) & -0.7127(29) \\
     & SPS & -0.7146(21) & -0.7133(17) & -0.7129(12) & -0.7128(13) \\
\hline
0.28 & HMC & -0.7811(31) & -0.7783(58) & -0.776(11) & -0.775(17) \\
     & SPS & -0.7811(24) & -0.7784(16) & -0.7784(47) & -0.7760(21) \\
\hline
0.29 & HMC & -0.8963(39) & -0.8920(63) & -0.889(12) & -0.891(18) \\
     & SPS & -0.8956(21) & -0.8932(34) & -0.8931(50) & -0.8925(31) \\
\hline
0.30 & HMC & -1.0545(48) & -1.0500(73) & -1.047(15) & -1.049(20) \\
     & SPS & -1.0544(33) & -1.0516(24) & -1.0502(38) & -1.0493(21) \\

\bottomrule
\end{tabular}
}
\caption{
Free energy density as a function of $\kappa$ for lattice sizes $L=16,32,48,64$. Results from HMC and SPS are shown.
}
\label{tab:free_energy}
\end{table}

To obtain accurate estimates of the free energy density, we used HMC with 160,000 samples and SPS with 10,240 samples. The SPS free energy density is computed using Eq.~\eqref{eq:free_energy_importance}. The results from HMC and SPS are shown in Fig.~\ref{fig:free} and in Table~\ref{tab:free_energy}. We observe that the free energy density obtained by SPS agrees with the one from HMC within the error.

\subsection{Acceptance Rate and Training cost}

\begin{figure}[htbp]
\centering
\subfloat[$L=16$]{
\label{fig:Acceptance16}
\includegraphics[width=0.49\linewidth]{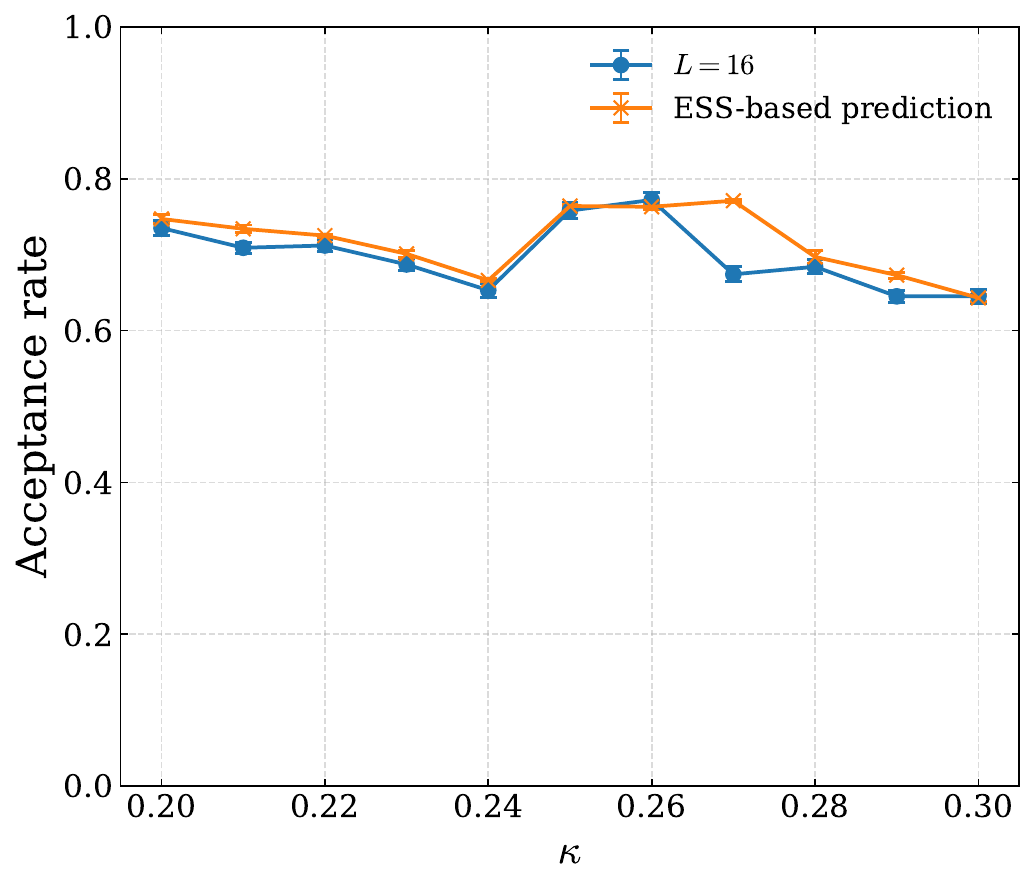}}
\subfloat[$L=32$]{
\label{fig:Acceptance32}
\includegraphics[width=0.49\linewidth]{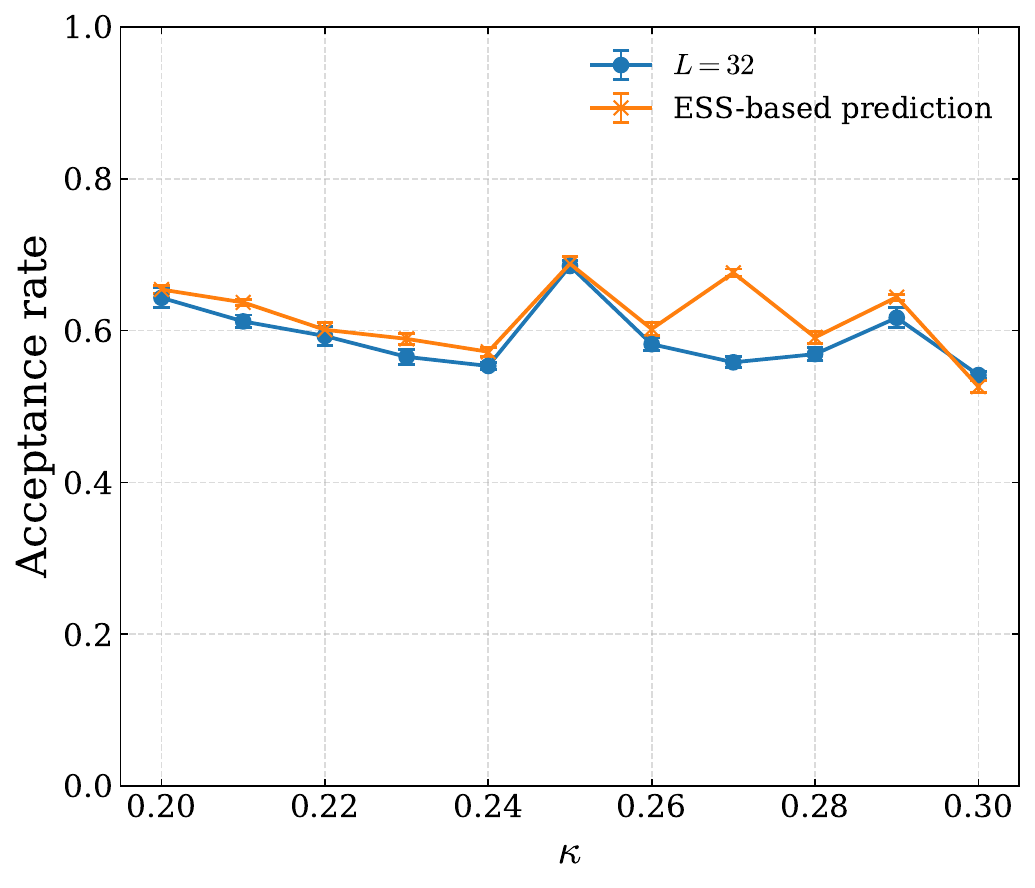}}\\
\subfloat[$L=48$]{
\label{fig:Acceptance48}
\includegraphics[width=0.49\linewidth]{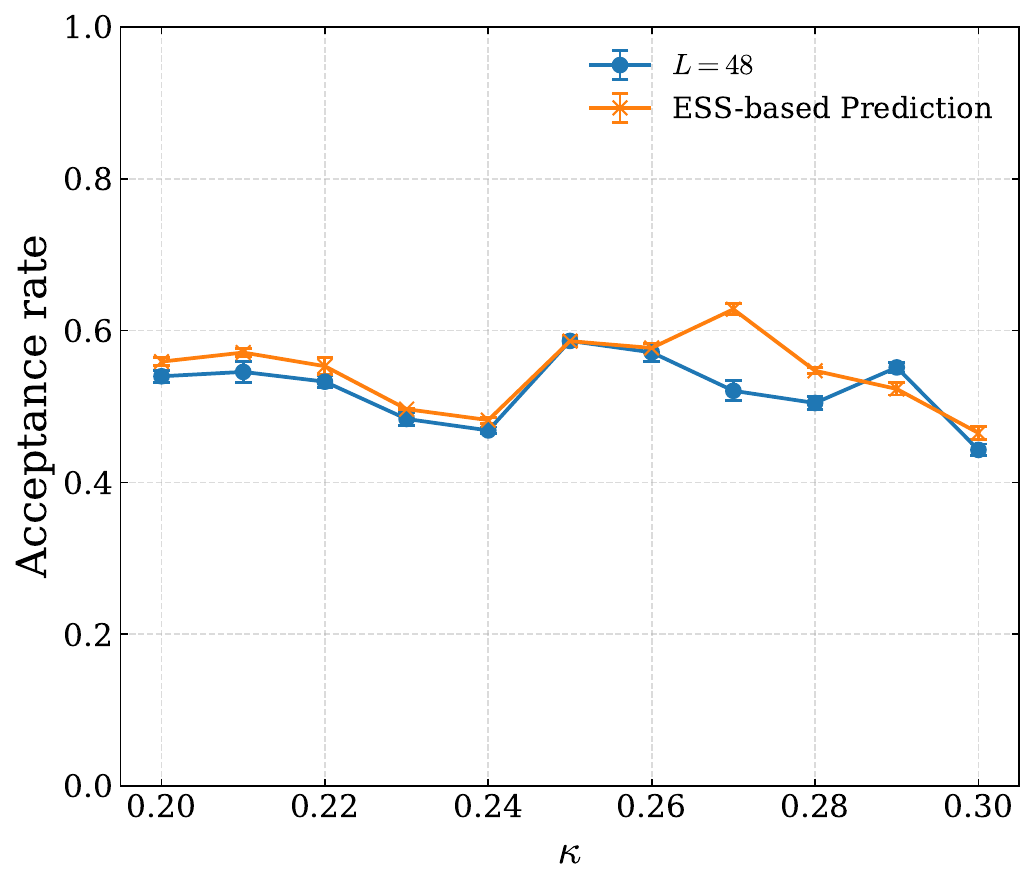}}
\subfloat[$L=64$]{
\label{fig:Acceptance64}
\includegraphics[width=0.49\linewidth]{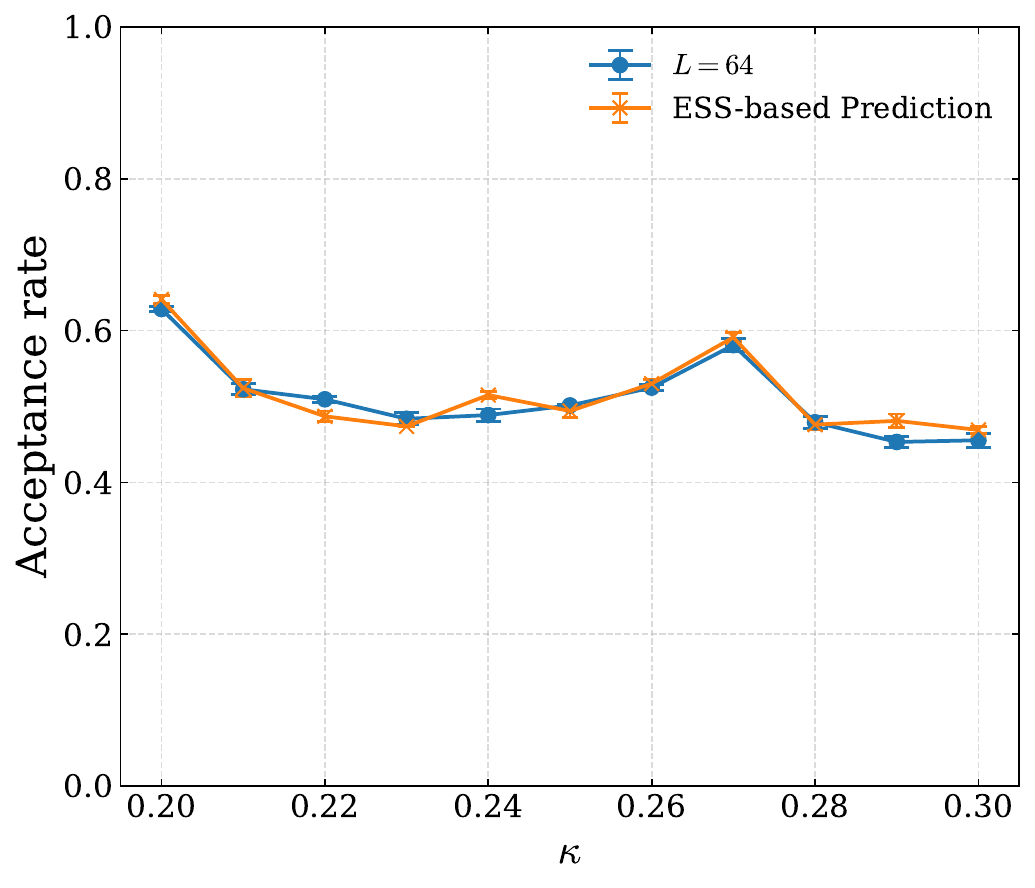}}
\caption{Acceptance rate from SPS with 4096 samples on four different lattice volumes. The blue circles show the directly measured IMH acceptance rate of the SPS, while the orange crosses show the corresponding ESS-based prediction from Eq.~\eqref{eq:ess_based_prediction}.}
\label{fig:Acceptance}
\end{figure}
In Fig.~\ref{fig:Acceptance}, we investigate the dependence of the Independence Metropolis--Hastings (IMH) acceptance rate on the hopping parameter $\kappa$ and the system size $L$. The acceptance rate is computed from a Markov chain comprising 4096 samples generated by the SPS, with the chain's structure and transition kernels detailed in Fig.~\ref{fig:Metropolis}.
The ESS-based prediction is given by
\begin{equation}
\langle A \rangle
\simeq \mathrm{erfc}\left(
\frac{1}{2}\sqrt{-\ln\left(\mathrm{ESS}/N_{\mathrm{s}}\right)}
\right), \qquad
\frac{\mathrm{ESS}}{N_{\mathrm{s}}}
\simeq
\frac{\langle w(\phi;\tau)\rangle^2}{\langle w^2(\phi;\tau)\rangle},
\label{eq:ess_based_prediction}
\end{equation}
where $\mathrm{ESS}/N_{\mathrm{s}}$ is the normalized effective sample size. The volume scaling of the $\mathrm{ESS}/N_{\mathrm{s}}$  and its relation to the fluctuations of the importance weights are discussed in detail in Appendix~\ref{app:ess_generation_depth}.

\begin{table}[htbp]
\centering
\scalebox{0.98}{
\setlength{\tabcolsep}{8pt}
\begin{tabular}{
c c
S[table-format=1.4(2)]
S[table-format=1.4(2)]
S[table-format=1.4(2)]
S[table-format=1.4(2)]
}
\toprule
{$\kappa$} & {Method}
& {$L=16$}
& {$L=32$}
& {$L=48$}
& {$L=64$} \\
\midrule

0.20 & {Prediction} & 0.747(6) & 0.654(5) & 0.559(5) & 0.641(5) \\
     & {SPS}        & 0.735(10) & 0.643(13) & 0.540(8) & 0.628(3) \\
\hline

0.21 & {Prediction} & 0.734(5) & 0.637(4) & 0.571(5) & 0.524(11) \\
     & {SPS}        & 0.709(7) & 0.612(8) & 0.546(14) & 0.523(7) \\
\hline

0.22 & {Prediction} & 0.725(2) & 0.601(9) & 0.553(11) & 0.487(7) \\
     & {SPS}        & 0.712(8) & 0.593(12) & 0.533(8) & 0.489(4) \\
\hline

0.23 & {Prediction} & 0.701(5) & 0.589(7) & 0.496(4) & 0.473(7) \\
     & {SPS}        & 0.687(8) & 0.565(10) & 0.484(9) & 0.504(8) \\
\hline

0.24 & {Prediction} & 0.666(2) & 0.572(6) & 0.482(4) & 0.515(5) \\
     & {SPS}        & 0.653(9) & 0.553(5) & 0.468(5) & 0.489(8) \\
\hline

0.25 & {Prediction} & 0.764(2) & 0.688(9) & 0.586(6) & 0.494(8) \\
     & {SPS}        & 0.758(11) & 0.685(7) & 0.586(8) & 0.501(2) \\
\hline

0.26 & {Prediction} & 0.763(4) & 0.602(9) & 0.577(6) & 0.530(6) \\
     & {SPS}        & 0.772(10) & 0.582(8) & 0.571(12) & 0.525(4) \\
\hline

0.27 & {Prediction} & 0.771(2) & 0.676(5) & 0.628(7) & 0.591(6) \\
     & {SPS}        & 0.674(10) & 0.558(7) & 0.521(13) & 0.581(8) \\
\hline

0.28 & {Prediction} & 0.697(9) & 0.591(8) & 0.547(4) & 0.476(7) \\
     & {SPS}        & 0.684(9) & 0.569(8) & 0.505(8) & 0.479(8) \\
\hline

0.29 & {Prediction} & 0.673(4) & 0.644(4) & 0.523(8) & 0.481(9) \\
     & {SPS}        & 0.645(8) & 0.617(13) & 0.551(7) & 0.443(7) \\
\hline

0.30 & {Prediction} & 0.643(5) & 0.526(8) & 0.465(8) & 0.469(5) \\
     & {SPS}        & 0.645(9) & 0.541(5) & 0.443(8) & 0.455(9) \\

\bottomrule
\end{tabular}
}
\caption{Acceptance rates as a function of the hopping parameter
$\kappa$ for lattice sizes $L=16,32,48,64$. The predictions are given by Eq.~\eqref{eq:ess_based_prediction}}.
\label{tab:acc-rate}
\end{table}

As shown in Fig.~\ref{fig:Acceptance} and Table~\ref{tab:acc-rate}, the SPS proposal maintains a reasonably high IMH acceptance rate over the whole range of $\kappa$ considered. A clear volume dependence is nevertheless observed, with the acceptance rate generally decreasing as the lattice size increases. For the smallest lattice, $L=16$, the acceptance rate remains around $0.60$--$0.76$, with only mild fluctuations as $\kappa$ varies. For the largest lattice, $L=64$, the acceptance rate remains in the range $0.45$--$0.62$, which is still moderate, although it shows a slightly decreasing tendency as $\kappa$ increases. Deeper in the broken phase ($\kappa \gtrsim 0.29$), as $\kappa$ increases, the bimodal structure of the magnetization distribution becomes more pronounced with a larger separation between the two modes and sharper individual peaks. This indicates that the probability weight is increasingly concentrated around the two symmetry-related vacua, making the target distribution more structured and more difficult for the SPS proposal to approximate accurately.

The theoretical prediction obtained from Eq.~\eqref{eq:ess_based_prediction} shows overall reasonable agreement with the directly measured IMH acceptance rate. The largest discrepancy is observed around $\kappa=0.27$, where the system is in the critical region and the fluctuations are enhanced. Since the theoretical estimate depends on the first and second moments of the importance weights through the finite-sample estimate of $\mathrm{ESS}/N_{\mathrm{s}}$, it is particularly sensitive to statistical fluctuations in the weight distribution. We therefore attribute the remaining difference mainly to the limited statistics used to estimate the ESS, rather than to a systematic failure of the predicted acceptance relation.

Nevertheless, in the pseudocritical region around $\kappa \simeq 0.26$--$0.28$, the SPS model maintains an acceptance rate comparable to the one observed in the symmetric phase. This demonstrates the high sampling efficiency of SPS near the pseudocritical point, despite the presence of enhanced fluctuations and long-range correlations.

\begin{figure}[htbp]
\centering
\subfloat[KL divergence]{
\label{fig:V_weight}
\includegraphics[width=0.49\linewidth]{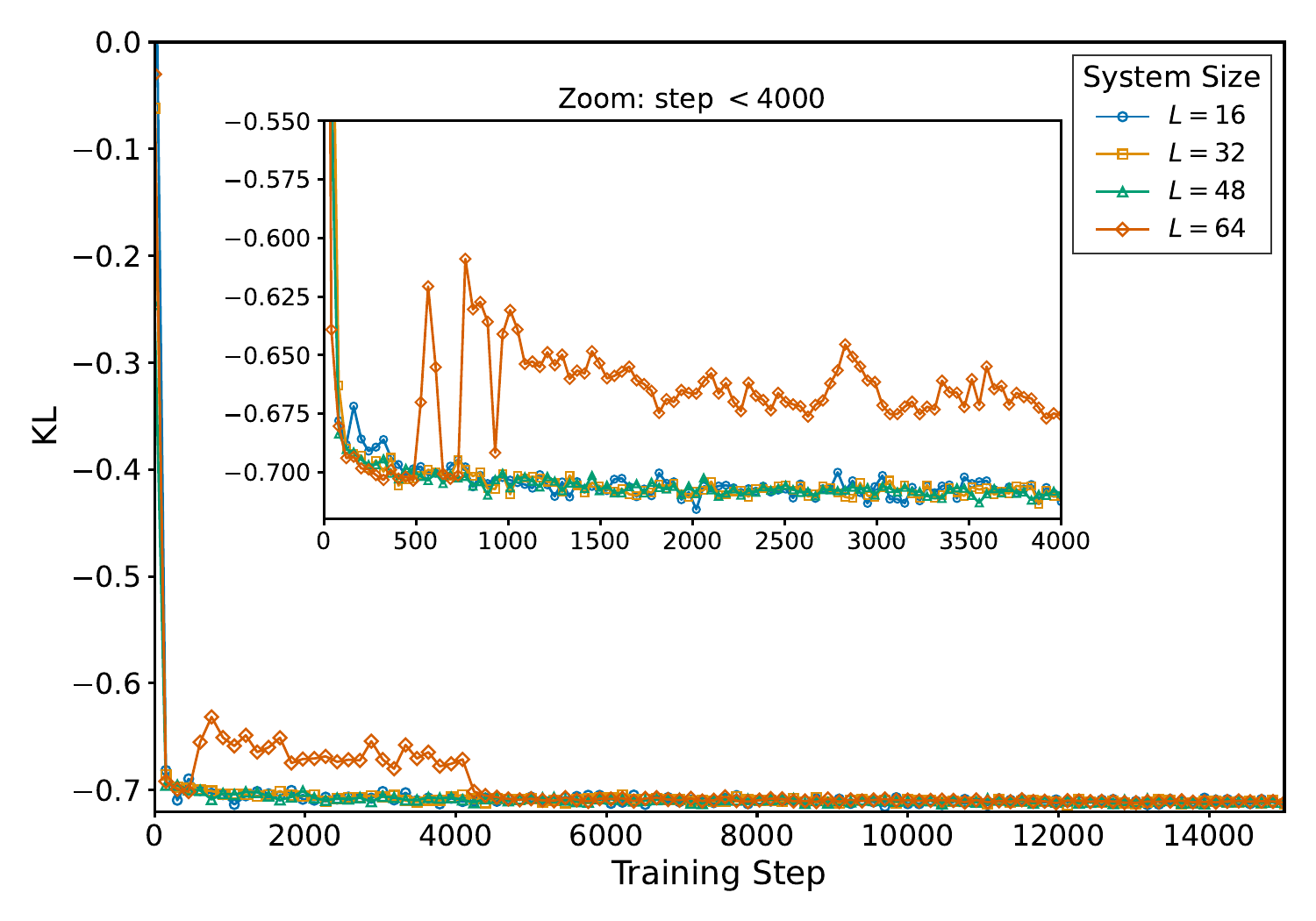}}
\subfloat[Variance of the log-weights]{
\label{fig:V_std}
\includegraphics[width=0.49\linewidth]{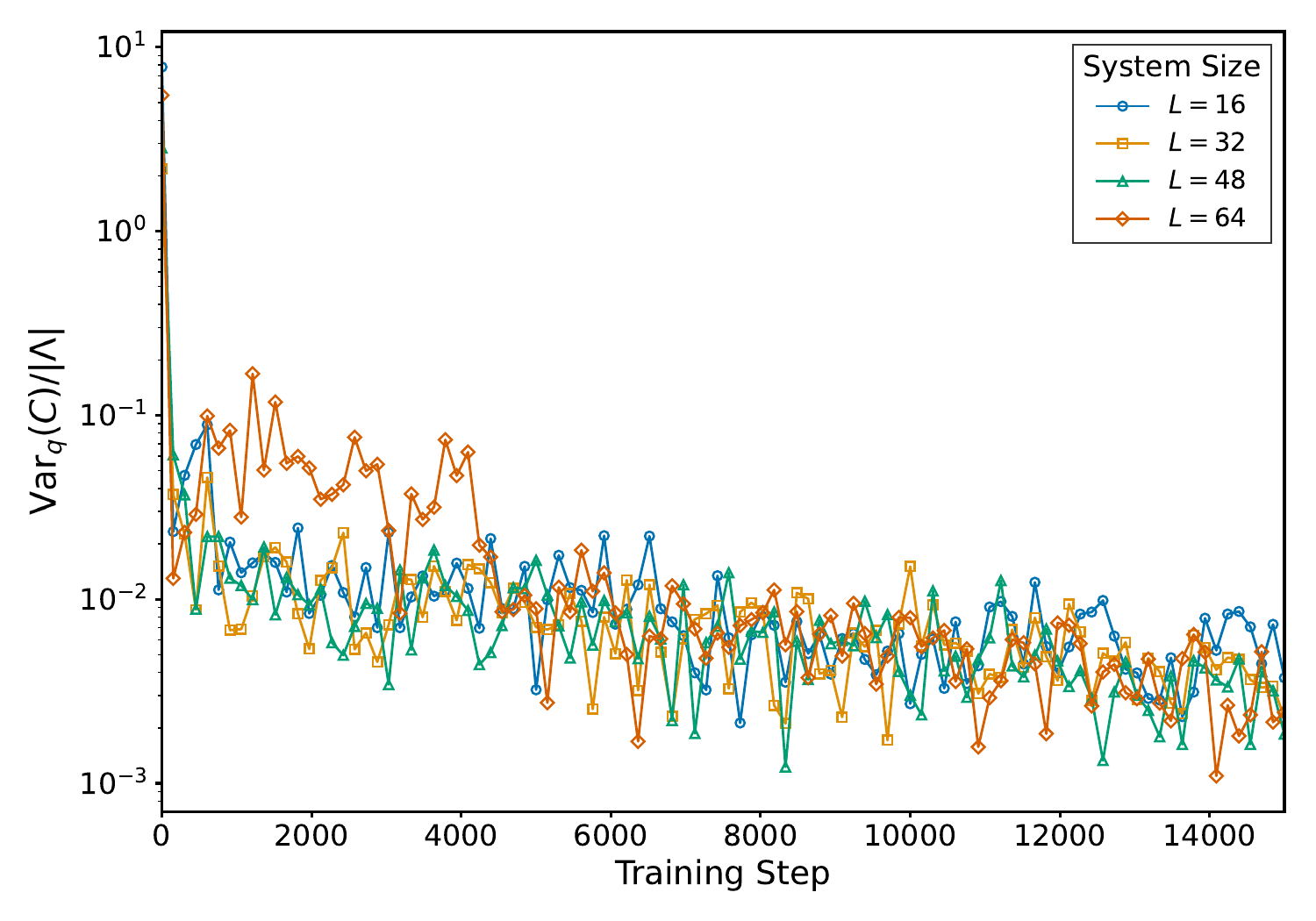}}
\caption{Training history at $\kappa=0.27$, close to the pseudocritical point. The markers and line colors correspond to different lattice sizes: $L=16$ (orange circles), $L=32$ (green squares), $L=48$ (red diamonds), and $L=64$ (blue triangles). (a) Evolution of the KL divergence (Eq.~\eqref{eq:kl_unnormalized}) with training steps. (b) Variance of the log-weights per site, Var$_q(C)/|\Lambda|$, versus training steps. The y-axis in panel (b) is on a logarithmic scale.}
\label{fig:History}
\end{figure}

Next, we investigate the training cost of the network by calculating the free energy and the variance of the corresponding $\log$ weights, with
\begin{equation}
     C = \log \tilde{\pi}^\ast(\phi) - \log q_\theta(\phi;\tau)
\end{equation}
at $\kappa=0.27$ across different training steps. We focus on this specific value of the hopping parameter, as in the pseudocritical region more training steps are required for the network to effectively learn the long-range correlations. The calculated numerical values and errors are presented in Fig.~\ref{fig:History}.  In Fig.~\ref{fig:V_weight}, we display the evolution of the KL divergence between the forward and unnormalized backward path measure (Eq.~\eqref{eq:kl_unnormalized}) as a function of training iterations for various lattice sizes. For all lattice sizes, the KL divergence exhibits a steep initial descent within the first 5,000 steps, rapidly converging to the target value, and subsequently stabilizing at approximately $-0.714$ for the remaining steps.  Fig.~\ref{fig:V_std} illustrates Var$_q(C)$ versus the training steps. A lower variance implies a smaller discrepancy between the SPS model distribution and the target distribution, which correlates with a higher acceptance rate. We observe a trend similar to that in Fig.~\ref{fig:V_weight}, in which the variance undergoes a sharp decline during the first 1,000 steps. Different from Fig.~\ref{fig:V_weight}, it is followed by a gradual decrease, before finally converging to a constant value.

\begin{table}[htbp]
    \centering
    \begin{tabular}{lcccc}
        \toprule
        $L$ & 16 & 32 & 48 & 64\\
        \midrule
        Generating time & $\sim 1.1$\,min & $\sim 1.5$\,min & $\sim 1.9$\,min&$\sim 2.1$\,min \\
        Training Time & ~$\sim 1.1$\,hours & ~$\sim 1.1$\,hour & ~$\sim 1.1 $\,hours &~$\sim 1.2 $\,hours\\
        \bottomrule
    \end{tabular}
     \caption{SPS performance for different lattice sizes at $\kappa=0.27$ and $\lambda=0.022$, including generation time ($T=$ 2,500 diffusion steps, 4096 samples generated), and training time ($T=$ 250 diffusion steps, 15,000 training step, batch size 12).}
     \label{tab:method_comparison}
\end{table} 

Table~\ref{tab:method_comparison} summarizes the computational cost of SPS for different lattice sizes at $\kappa=0.27$ and $\lambda=0.022$. The generation time is measured using $T=2500$ diffusion steps, while the training time corresponds to $T=250$ diffusion steps. As the lattice size increases from $L=16$ to $L=64$, the generation time increases moderately from about $1.1$ minutes to about $2.1$ minutes, whereas the training time remains nearly constant, around $1.1$--$1.2$ hours. Details of the computational platform are provided in Appendix~\ref{app:network_training}.

\subsection{Autocorrelation Time}
The autocorrelation function serves as a crucial observable for investigating critical slowing down as $\kappa$ approaches the pseudocritical region. Critical slowing down, a hallmark challenge in conventional Markov chain Monte Carlo methods like HMC, manifests itself as a dramatic increase in the autocorrelation time near a phase transition, severely degrading sampling efficiency.
To quantify this effect and benchmark the performance of SPS, we compute the normalized autocorrelation function for the absolute magnetization $|M|$,

\begin{equation}
    \bar{C}_{|M|}(t) \equiv C_{|M|}(t)/C_{|M|}(0),
\end{equation}
where the unnormalized autocorrelation function is defined as
\begin{align}
\begin{aligned}
    C_{|M|}(t) & = \left\langle ({|M|}_{t_0} - \left\langle {|M|}_{t_0} \right\rangle)({|M|}_{t_0+t} - \left\langle {|M|}_{t_0+t} \right\rangle) \right\rangle 
    \\
        & = \left\langle {|M|}_{t_0}{|M|}_{t_0+t} \right\rangle - \left\langle {|M|}_{t_0} \right\rangle\left\langle {|M|}_{t_0+t} \right\rangle.
\end{aligned}
\end{align}
Here, for SPS, $t_0$ and $t$ denote the starting point and the lag along the Markov chain generated by the IMH algorithm, which is distinct from the diffusion time.

\begin{figure}[htbp]
\centering
\includegraphics[width=0.7\linewidth]{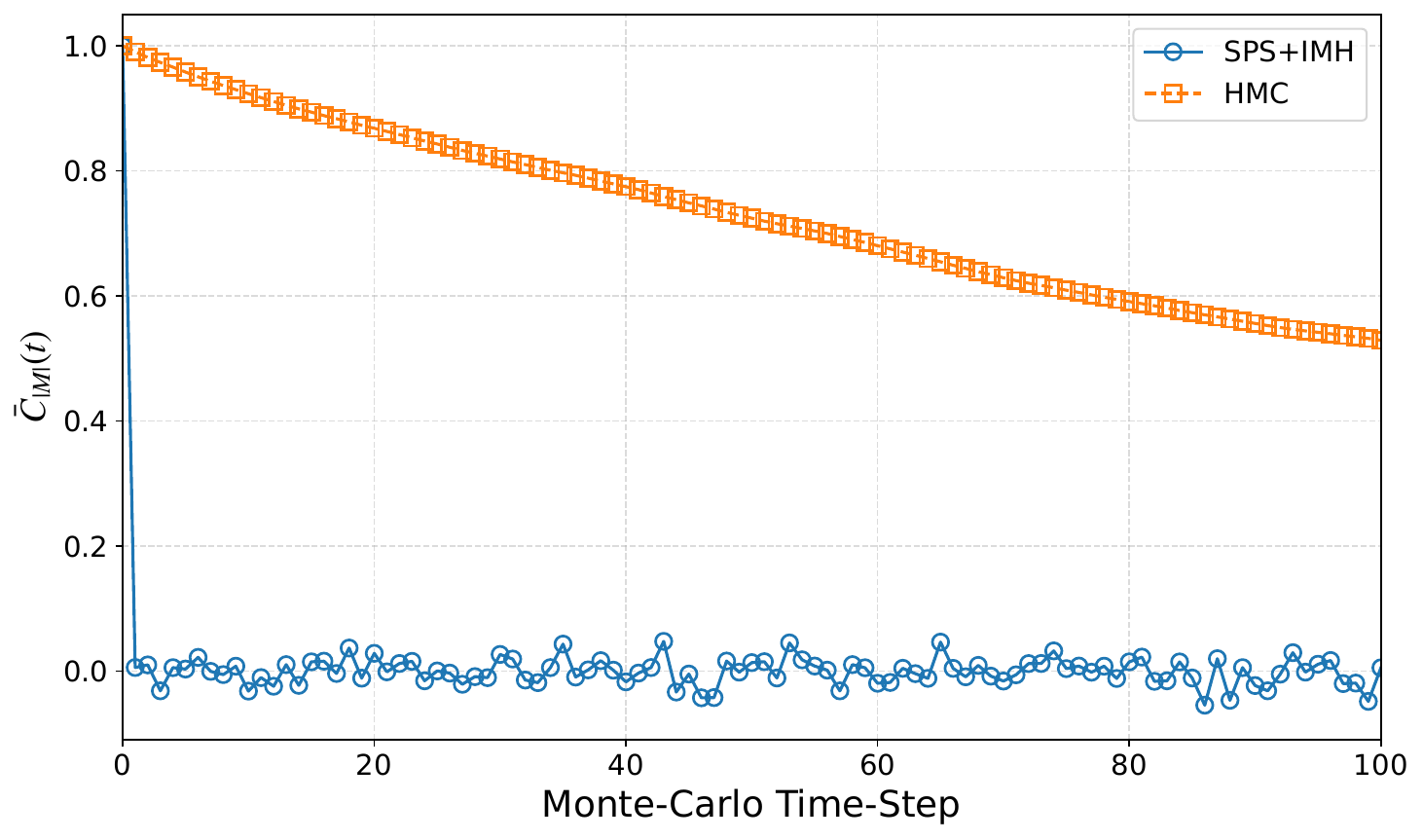}
\caption{Autocorrelation function for the absolute magnetization $|M|$ at $\kappa =0.27$ and $L=64$.  Blue circles are for SPS$+$IMH and yellow squares for HMC.}
\label{fig:Autocorrelation}
\end{figure}

We focus our analysis on $\kappa=0.27$, close to the pseudocritical point, for a lattice size of $L=64$, where critical slowing down is most pronounced. The results are presented in Fig.~\ref{fig:Autocorrelation}. As can be seen, the autocorrelation function in the case of SPS$+$IMH decays significantly faster than the one for HMC, demonstrating the efficiency of SPS in generating decorrelated samples near the pseudocritical point.

A key quantitative measure derived from the autocorrelation function is the integrated autocorrelation time, $\tau_{|M|,int}$, which estimates the number of MCMC steps required to generate an effectively independent sample. It is defined as
\begin{equation}
    \tau_{|M|,{\mathrm{int}}}={\frac{1}{2}}+{\frac{1}{C_{|M|}(0)}}\sum_{t=1}^{t_\text{max}}C_{|M|}(t),
\end{equation}
where the summation window \(t_{\rm max}\) is chosen using an automatic windowing procedure. Starting from \(t=1\), the sum is accumulated until either \(\rho_{|M|}(t)\le 0\) or the self-consistent window criterion
\begin{equation}
t > c\,\tau_{\rm int}(t)
\end{equation}
is satisfied, with \(c=6\). Here \(\tau_{\rm int}(t)\) denotes the partial integrated autocorrelation time evaluated up to lag \(t\). This procedure avoids introducing a fixed, manually chosen cutoff in the tail of the autocorrelation function.

The computed integrated autocorrelation times are $\tau^{\text{SPS$+$IMH}}_{|M|,int}\simeq 0.5$ for SPS with IMH correction and $\tau^{\text{HMC}}_{|M|,int}\simeq 160$ for HMC.  This substantial reduction highlights the effectiveness of SPS in mitigating the critical slowing down problem. We note that these autocorrelation times are expressed in different units---IMH steps for SPS$+$IMH and trajectories for HMC---and therefore do not directly translate into a cost-matched comparison of overall sampling efficiency, which would also need to account for the training and generation costs reported in Table~\ref{tab:method_comparison}.

\section{Conclusion}
\label{sec:4}

In this paper we proposed a neural network method based on nonequilibrium thermodynamics, called the Stochastic Path Sampler (SPS).
It employs two independent neural networks to learn the discretized Langevin dynamics paths of the forward process (from the prior distribution to the target distribution) and the backward process (from the target distribution back to the prior distribution), respectively, together with a third small network that parameterizes the learnable diffusion coefficient $\sigma_\theta(t)$. The method defines the logarithm of the ratio of the forward and backward path probabilities over the entire trajectory as an entropy, and trains the networks by minimizing the KL divergence in path space. This process aims to globally reduce the irreversibility of the sampling process, ensuring that the final configuration generated by the forward process approximates the target (Boltzmann) distribution. To guarantee sampling accuracy, an Independence Metropolis--Hastings correction step based on the full trajectory is introduced.

To validate the effectiveness of SPS, we systematically benchmark it for the two-dimensional lattice $\phi^4$ scalar field model and comprehensively compare it with the standard Hybrid Monte Carlo (HMC) algorithm. The study focuses on key physical observables such as magnetization, susceptibility, and free energy density. The results show that SPS with IMH correction reproduces all physical quantities computed by HMC with high accuracy, including their behavior in the symmetric phase, the pseudocritical region, and the symmetry‑broken phase. Moreover, in the pseudocritical region ($\kappa$ = 0.27), the integrated autocorrelation time of the absolute magnetization for SPS$+$IMH is significantly lower than that for HMC, $\tau^{\text{SPS$+$IMH}}_{|M|,int}\simeq 0.5$ versus $\tau^{\text{HMC}}_{|M|,int}\simeq 160$ respectively, measured in units of IMH steps and HMC trajectories.

We have also studied the scaling of the algorithmic cost and of the acceptance rate with the volume. At a fixed number of training steps, the cost of the algorithm shows only a mild dependence on the volume across the sizes studied. While the acceptance rate of the IMH step is found to decrease with the volume, the algorithm remains efficient up to the largest investigated size, $L=64$. We expect that decreasing the discretisation step $\Delta t$ of the Langevin dynamics will increase the acceptance. A detailed study of the scaling of the algorithmic cost with $\Delta t$ is outside the scope of the present work, and will be pursued elsewhere.

A natural next step is to extend SPS from real scalar fields to lattice gauge theories, where the degrees of freedom are group-valued. Two-dimensional compact $U(1)$ gauge theory provides a natural first benchmark, since it would allow us to test whether the stochastic path construction can alleviate topological freezing relative to standard sampling methods. A second important direction is to investigate the scalability of SPS to larger volumes and higher spacetime dimensions. While the present results in two-dimensional $\phi^4$ theory are promising, applications to three- and four-dimensional systems will require more expressive architectures and a systematic study of how the acceptance rate, training cost, and autocorrelation time scale with system size. In particular, the convolution kernels of the present architecture grow with the lattice extent $L$ (see Appendix~\ref{sec:Network}), so that the network is effectively global and its parameter count increases with the volume; more scalable, local architectures will need to be explored for higher-dimensional applications.

\el
\noindent
\Needspace{12\baselineskip}
\noindent{\bf Acknowledgements} --
We thank Lingxiao Wang for discussions.
This project is supported in part by the Royal Society International Exchanges 2024 Global Round 2 (IES\textbackslash R2\textbackslash 242215).
SC, GA, and BL thank KZ for hospitality at CUHK-Shenzhen, and KZ thanks BL and GA for hospitality at QMUL and SU.
SC is supported by the China Scholarship Council (No.~202308420042) and the Swansea University joint PhD project.
GA is supported by STFC grant ST/X000648/1 and by a Royal Society Leverhulme Trust Senior Research Fellowship.
BL is supported by STFC Consolidated Grants ST/X000648/1 and ST/X00063X/1.
KZ is supported by the CUHK-Shenzhen University Development Fund under grant Nos.~UDF01003041 and UDF03003041, the Ministry of Science and Technology of China under grant No.~2024YFA1611004, the National Natural Science Foundation of China under grant No.~92570117, the Shenzhen Peacock Fund under grant No.~2023TC0179, and the Shenzhen General Fund under grant No.~JCYJ20250604141001001.
We acknowledge the support of the Supercomputing Wales and AccelerateAI projects, which are part-funded by the European Regional Development Fund (ERDF) via the Welsh Government.

\noindent
{\bf Research Data and Code Access}  --
The code and data used in this manuscript are available at \url{https://github.com/Chern2/Stochastic-Path-Sampler-For-Lattice-Field-Theory}.

\noindent
{\bf Open Access Statement} -- For the purpose of open access, the authors have applied a Creative Commons Attribution (CC BY) licence to any Author Accepted Manuscript version arising.

\noindent
{\bf Use of AI Statement} -- The authors used Claude (Anthropic) and ChatGPT (OpenAI) to support the literature review and to refine the consistency of the notation.

\appendix

\section{SPS with Nelson's Duality Relation}
\label{app:duality_relation}

This section presents the relationship between the diffusion model and SPS.
\subsection{Path Probability Ratio Under Gaussian Transition Kernels}

Under the isotropic diffusion assumption, the forward kernel is
\begin{equation}
q_{\mathrm{F}}(\bm{s}_{i+1}|\bm{s}_i)
\propto
\exp\Big(
-\frac{\|\bm{s}_{i+1}-\bm{s}_i-\sigma^2_{\theta}(t_{i})\bm{K}_{\theta,\mathrm{F}}(\bm{s}_i,t_{i}) \Delta t\|^2}{2\sigma^2_{\theta}(t_{i} )\Delta t}
\Big),
\end{equation}
and the backward kernel is

\begin{equation}
q_{\mathrm{B}}(\bm{s}_{i}|\bm{s}_{i+1})
\propto
\exp\Bigg(
-\frac{
\left\|
\bm{s}_{i}-\bm{s}_{i+1}
-\sigma^2_{\theta}\left(
t_i
\right)
\bm{K}_{\theta,\mathrm{B}}
\left(
\bm{s}_{i+1},
t_i
\right)
\Delta t
\right\|^2
}{
2\sigma_\theta^2\left(
t_i
\right)\Delta t
}
\Bigg).
\end{equation}

Denoting $\Delta\bm{s}_t = \bm{s}_{i+1}-\bm{s}_i$, the log ratio is

\begin{align}
&\log\frac{
q_{\mathrm{F}}(\bm{s}_{i+1}|\bm{s}_i)
}{
q_{\mathrm{B}}(\bm{s}_{i}|\bm{s}_{i+1})
}
=
-\frac{
\big(
\|\Delta\bm{s}_i
-\sigma_\theta^2(t_i)
\bm{K}_{\theta,\mathrm{F}}(\bm{s}_i,t_i)
\Delta t\|^2
-
\|-\Delta\bm{s}_i
-\sigma_\theta^2(t_i)
\bm{K}_{\theta,\mathrm{B}}
(\bm{s}_{i+1},
t_i)
\Delta t\|^2
\big)
}{
2\sigma_\theta^2(t_i)\Delta t
}
\\
&=
\big(
\bm{K}_{\theta,\mathrm{B}}
(\bm{s}_{i+1},
t_i)
+
\bm{K}_{\theta,\mathrm{F}}(\bm{s}_i,t_i)
\big)\cdot\Delta\bm{s}_i
\nonumber\\
&\quad+
\frac{1}{2}\sigma_\theta^2(t_i)\Delta t
\bigg(
\left\|
\bm{K}_{\theta,\mathrm{B}}
\left(\bm{s}_{i+1},
t_i
\right)
\right\|^2
-
\left\|
\bm{K}_{\theta,\mathrm{F}}
\left( \bm{s}_i,
t_i
\right)
\right\|^2
\bigg).
\end{align}

Furthermore, in the small step-size limit,
$\Delta\bm{s}_i\sim\mathcal{O}(\sqrt{\Delta t})$,
and the quadratic drift contribution is of order
$\mathcal{O}(\Delta t)$ per step.
Retaining only the leading term gives
\begin{equation}
\log\frac{
q_{\mathrm{F}}(\bm{s}_{i+1}|\bm{s}_i)
}{
q_{\mathrm{B}}(\bm{s}_{i}|\bm{s}_{i+1})
}
\approx
\bigg(
\bm{K}_{\theta,\mathrm{B}}
\left(
\bm{s}_{i+1},
t_i
\right)
+
\bm{K}_{\theta,\mathrm{F}}(\bm{s}_i,t_i)
\bigg)
\cdot\Delta\bm{s}_i.
\end{equation}

Summing over the entire trajectory yields
\[
\sum_{i=0}^{T-1}
\bigg(
\bm{K}_{\theta,\mathrm{B}}
\left(
\bm{s}_{i+1}, t_i
\right)
+
\bm{K}_{\theta,\mathrm{F}}(\bm{s}_i,t_i)
\bigg)
\cdot(\bm{s}_{i+1}-\bm{s}_i)
\approx
\log\frac{
\prod_i q_{\mathrm{F}}(\bm{s}_{i+1}|\bm{s}_i)
}{
\prod_i q_{\mathrm{B}}(\bm{s}_i|\bm{s}_{i+1})
}.
\]

Combined with the prior $\pi_0(\bm{s}_0)$
and the target $\tilde{\pi}^\ast(\bm{s}_T)$,
we obtain the summation version of
Eq.~\eqref{eq:tb_ratio_main}.

\subsection{Continuous Time Limit}

In the continuous limit, where $\bm{s}_{t+1}-\bm{s}_t\simeq d\bm{s}$, the trajectory sum approximates a line integral,
\begin{equation}
\int d\bm{s}(t)\cdot
\big(\bm{K}_{\theta,\mathrm{B}}(\bm{s},t)
+ \bm{K}_{\theta,\mathrm{F}}(\bm{s},t)\big)
\approx \log\tilde{\pi}^\ast(\bm{s}_T)-\log\pi_0(\bm{s}_0) + \text{const},
\end{equation}
which is Eq.~\eqref{eq:line_integral_main} from the main text.
If we assume that at any time slice $t$, the marginal distribution is $P(\bm{s},t)$, and the forward/backward processes are its time-forward/backward processes, then the integral over any closed loop must be zero, yielding the local relation
\begin{equation}
\bm{K}_{\theta,\mathrm{F}}(\bm{s},t)
+ \bm{K}_{\theta,\mathrm{B}}(\bm{s},t)
= \nabla_{\bm{s}}\log P(\bm{s}, t),
\end{equation}
which is Nelson's duality relation. Therefore, reducing irreversibility and  minimizing Eq.~\eqref{eq:loss} can be viewed as a global approximation to this condition under finite step size and finite capacity, while in the continuous limit the two are completely consistent. Meanwhile, the pair $(\bm{K}_{\theta,\mathrm{F}},\bm{K}_{\theta,\mathrm{B}})$ forms a vector field that pushes the initial distribution to the target distribution. In score-based diffusion models, it is typically used as the theoretical foundation for constructing the backward generative process from the forward diffusion process. However, in the framework of this paper, SPS does not directly fit the score and does not impose this pointwise relation; rather, it learns drift pairs $(\bm{K}_{\theta,\mathrm{F}},\bm{K}_{\theta,\mathrm{B}})$ that reduce the path-space mismatch in a global variational sense.

\section{SPS Network Architecture, Training, and Generation}
\label{sec:Network}

\subsection{Network Architecture}
\label{sec:architecture}
We employ a time-conditional convolution neural network, to parameterize the forward and backward drifts of the diffusion process on a two-dimensional lattice. The forward and backward drift networks share the same architecture but have independent parameters (weights). The network takes as state $\bm{s}_i$ a lattice configuration $\mathbb{R}^{L\times 8}$ together with a diffusion time variable $t$, and outputs a drift for the forward/backward Langevin process. 

The diffusion time $t$ is embedded using a fixed Fourier feature mapping followed by a learnable linear projection. Specifically, the time embedding is constructed as
\begin{equation}
    \gamma(t) = \text{ReLU}\bigg(\text{Dense}([\sin(Wt),\cos(Wt)])\bigg),
\end{equation}
where $W\in \mathbf{R}^{64}$ is a fixed random vector and the total embedding dimension is 128. The same time embedding module is shared across all convolution blocks.

\label{fig:architecture}
\begin{figure}[htbp]
\centering
\includegraphics[width=1\linewidth]{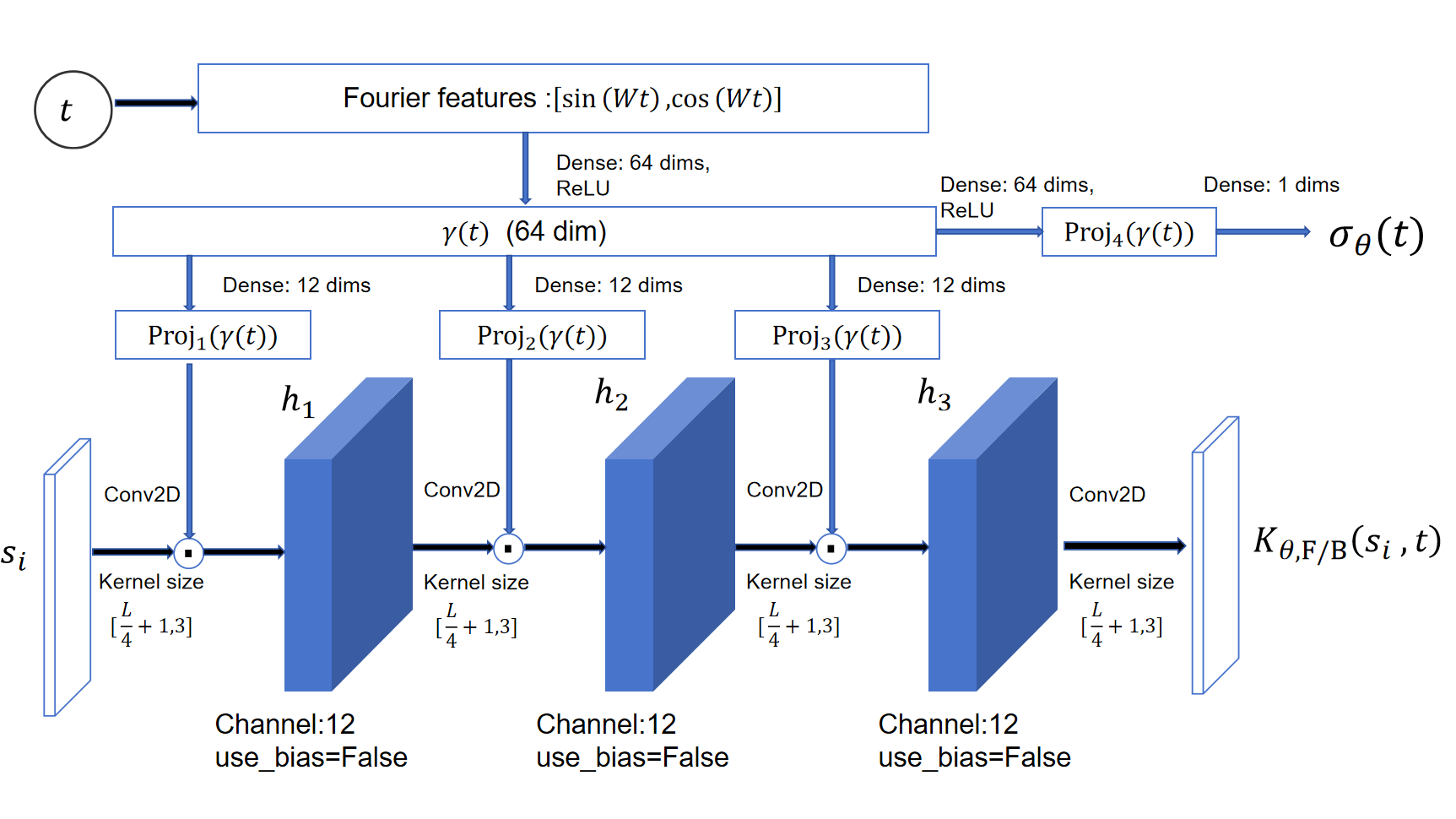}
\caption{Network architecture used to estimate the drift and diffusion coefficient for the forward and backward processes.}
\label{fig:network's architecture}
\end{figure}

The backbone consists of three convolution blocks with increasing dilation factors, followed by a final output layer. All convolution layers are implemented with circular padding in both spatial directions, ensuring exact periodic boundary conditions on the lattice.

The input state $\bm{s}$ is first processed by a cyclic padding convolution block with 12 output channels. Formally, the computation is given by
\begin{equation}
h_1 =
\tanh \bigg( \mathrm{Conv}_{[L/4+1, 3]} (\bm{s}_i) \odot \exp\left[\mathrm{Proj}_1(\gamma(t))\right] \bigg),
\end{equation}
where $\mathrm{Conv}_{[L/4+1, 3]}$  denotes a cyclic padding convolution with  filter size $[L/4+1,3]$ , strides $=1$, and $\odot$ denotes channel-wise multiplication after broadcasting the projected time embedding over the spatial lattice. The term $\mathrm{Proj}_1(\gamma(t))$ corresponds to a projection of the time embedding, which is broadcast and multiplied to all spatial locations.

The second block applies another cyclic convolution with kernel size $[L/4+1,3]$ to $h_1$, followed by the same type of time-dependent multiplicative modulation. A residual connection from $h_1$ is then added: 
\begin{equation}
h_2=\frac{1}{2} \tanh \bigg( \mathrm{Conv}_{[L/4+1, 3]} (h_1) \odot \mathrm{Proj}_2(\gamma(t)) \bigg) + h_1.
\end{equation}
The term $\mathrm{Proj}_2(\gamma(t))$ corresponds to a projection of the time embedding.

The third block further processes $h_2$ using a cyclic convolution with kernel size $[L/4+1, 3]$. The time-conditioned update is again multiplicative, and the output is combined with $h_2$ through a residual connection:
\begin{equation}
h_3=\tanh \bigg(\mathrm{Conv}_{[L/4+1, 3]} (h_2) \odot \mathrm{Proj}_3(\gamma(t))\bigg) + h_2.
\end{equation}
The term $\mathrm{Proj}_3(\gamma(t))$ corresponds to a projection of the time embedding.

The final drift field is produced by applying a cyclic padding convolution with kernel size $[L/4+1, 3]$ to the output of the third block,
\begin{equation}
\bm{K}_{\theta, \mathrm{F}/\mathrm{B}}(\bm{s}_i,t) =\mathrm{Conv}_{[L/4+1, 3]}(h_3),
\end{equation}
yielding a single-channel output defined on the same spatial lattice as the input. To respect the \(Z_2\) symmetry of the lattice \(\phi^4\) theory, the drift networks are constructed to be \(Z_2\)-equivariant under the global field transformation
\(\phi \to -\phi\). Namely, the learned drift satisfies
\begin{equation}
\bm{K}_{\theta,\mathrm{F/B}}(-\bm{\phi},t)
=
-\bm{K}_{\theta,\mathrm{F/B}}(\bm{\phi},t).
\end{equation}
As a result, the stochastic dynamics does not explicitly favor either of the two symmetry-related sectors in the broken phase. This helps the learned proposal maintain support over both \(Z_2\)-related modes and reduces the risk of mode collapse onto a single sector.

The convolution kernels in these four convolutional layers are $L$-dependent, which allows the network to construct the correct global correlations for different lattice sizes. 

In addition to imposing exact periodic boundary conditions, we incorporate the local point-group symmetry of the lattice by explicitly symmetrizing the convolution kernels over all shape-preserving transformations of their spatial support. For a square kernel, the effective kernel is constructed by averaging over the eight elements of the dihedral group \(D_4\),
\begin{equation}
    W_{\mathrm{eff}}
    =
    \frac{1}{|D_4|}
    \sum_{g\in D_4}
    T_g W,
\end{equation}
where \(T_g\) denotes the action of the corresponding rotation or reflection on the spatial indices of the kernel.

For a rectangular kernel, rotations by \(90^\circ\) or \(270^\circ\) do not preserve the kernel shape. Therefore, the symmetrization is restricted to the four-element \(D_2\) subgroup,
\begin{equation}
    W_{\mathrm{eff}}
    =
    \frac{1}{|D_2|}
    \sum_{g\in D_2}
    T_g W,
\end{equation}
which consists of the identity transformation, reflections along the two lattice axes, and a \(180^\circ\) rotation. By construction, the effective kernel satisfies
\begin{equation}
    T_g W_{\mathrm{eff}}
    =
    W_{\mathrm{eff}},
    \qquad
    g\in D_4
    \ \text{or}\
    D_2.
\end{equation}
Consequently, the corresponding convolution operator is equivariant under the supported local lattice transformations,
\begin{equation}
    \mathrm{Conv}_{W_{\mathrm{eff}}}
    \left(T_g \phi\right)
    =
    T_g
    \mathrm{Conv}_{W_{\mathrm{eff}}}
    \left(\phi\right).
\end{equation}
This construction prevents the network from assigning independent parameters to symmetry-equivalent local orientations and ensures that the learned drift respects the local rotational and reflection symmetries compatible with the geometry of the convolution kernel.

We also learn the scalar diffusion coefficient. We parameterize the time-dependent diffusion coefficient by a small neural network,
\begin{equation}
    \sigma_{\theta}(t)
    =
    \mathrm{Sigmoid}
    \left[
    \mathrm{Dense}
    \left(
    \mathrm{ReLU}
    \left(
    \mathrm{Dense}(\gamma(t))
    \right)
    \right)
    \right],
    \label{eq:sigma_network}
\end{equation}
which guarantees $\sigma_{\theta}(t)>0$. In our implementation, the same scalar diffusion schedule is used for the forward and backward kernels,
\begin{equation}
    \sigma_{\mathrm{F}}(t_i)
    =
    \sigma_{\mathrm{B}}(t_{i+1})
    \equiv
    \sigma_{\theta}(t_i).
    \label{eq:sigma_shared}
\end{equation}
The diffusion coefficient is optimized jointly with the forward and backward drift networks.

\subsection{Network Training}
\label{app:network_training}

The SPS networks are trained by minimizing the path-space KL divergence between the learned forward path measure and the unnormalized backward path measure. During training, an initial configuration is first sampled from the Gaussian prior,
\begin{equation}
    \mathbf{s}_0 \sim \pi_0(\mathbf{s}) = \mathcal{N}(0,I).
\end{equation}
The configuration is then evolved through the learned forward stochastic dynamics,
\begin{equation}
    \mathbf{s}_{i+1} = \mathbf{s}_i + \sigma_{\theta}^{2}(t_i) K_{\theta,\mathrm{F}}(\mathbf{s}_i,t_i)\Delta t + \sigma_{\theta}(t_i)\sqrt{\Delta t}\,\xi_i,
    \qquad
    \xi_i \sim \mathcal{N}(0,I),
    \label{eq:training_forward_update}
\end{equation}
for $i=0,\ldots,T-1$. The final configuration $\mathbf{s}_{T}$ is identified with the generated field
configuration $\phi$.

At each time step, the forward transition probability is given by the Gaussian kernel
\begin{equation}
    q_{\mathrm{F}}(\mathbf{s}_{i+1}|\mathbf{s}_i) = \mathcal{N}\left(\mathbf{s}_{i+1}; \mathbf{s}_i+\sigma_{\theta}^{2}(t_i)K_{\theta,\mathrm{F}}(\mathbf{s}_i,t_i)\Delta t,
    \sigma_{\theta}^{2}(t_i)\Delta t\,I
    \right).
    \label{eq:qF_training}
\end{equation}
For the same pair of consecutive states, the backward kernel is evaluated as

\begin{equation}
    q_{\mathrm{B}}(\mathbf{s}_i|\mathbf{s}_{i+1})
    = \mathcal{N}\left(
    \mathbf{s}_i;
    \mathbf{s}_{i+1}
    + \sigma_{\theta}^{2}(t_i)
    K_{\theta,\mathrm{B}}
    \left(
        \mathbf{s}_{i+1}, t_i \right)\Delta t,
    \sigma_{\theta}^{2}(t_i)\Delta t\,I
    \right).
    \label{eq:qB_training}
\end{equation}

Here $K_{\theta,\mathrm{F}}$ and $K_{\theta,\mathrm{B}}$ are the forward and backward drift networks, respectively, and $\sigma_{\theta}(t)$ is the learned scalar diffusion coefficient.

The log density of the SPS proposal is accumulated along the trajectory as
\begin{equation}
    \log q_{\theta}(\mathbf{s}_{T};\tau)
    =
    \log \pi_0(\mathbf{s}_0)
    +
    \sum_{i=0}^{T-1}
    \left[
    \log q_{\mathrm{F}}(\mathbf{s}_{i+1}|\mathbf{s}_i)
    -
    \log q_{\mathrm{B}}(\mathbf{s}_i|\mathbf{s}_{i+1})
    \right].
    \label{eq:training_logq}
\end{equation}
The training objective is then defined by
\begin{equation}
    \mathcal{L}_{\mathrm{SPS}}(\theta)
    =
    \mathbb{E}_{q_{\mathrm{F}}}
    \left[
    \log q_{\theta}(\mathbf{s}_{T};\tau)
    -
    \log \tilde{\pi}^{*}(\mathbf{s}_{T})
    \right],
    \label{eq:sps_training_loss}
\end{equation}
where
\begin{equation}
    \log \tilde{\pi}^{*}(\mathbf{s}_{T}) = -S(\mathbf{s}_{T})
\end{equation}
is the unnormalized log target density. Since the partition function of the target distribution is not required in Eq.~\eqref{eq:sps_training_loss}, the SPS can be trained without reference samples from HMC or any other Markov chain sampler.

The expectation value in Eq.~\eqref{eq:sps_training_loss} is estimated with mini-batches of stochastic trajectories. Gradients are backpropagated through the full discretized trajectory, including the forward drift network, the backward drift network, and the diffusion network. The stochastic noise variables $\{\xi_i\}$ are sampled independently at each time step and for each element of the mini-batch. In practice, the loss is estimated as
\begin{equation}
    \widehat{\mathcal{L}}_{\mathrm{SPS}}(\theta)
    =
    \frac{1}{N_s}
    \sum_{n=1}^{N_s}
    \left[
    \log q_{\theta}(\mathbf{s}_{T}^{(n)};\tau)
    -
    \log \tilde{\pi}^{*}(\mathbf{s}_{T}^{(n)})
    \right],
    \label{eq:sps_minibatch_loss}
\end{equation}
where $N_s$ is the batch size.

In our numerical experiments, the diffusion interval $[0,1]$ is discretized into $T=250$ steps during training. The networks are optimized using Adam with an initial learning rate of $10^{-3}$. A delayed cosine decay schedule is applied after the first $5000$ optimization steps, reducing the learning rate to $5\times 10^{-5}$ over the following $10000$ steps. The total number of training iterations is $15000$, and the batch size is set to $N_s=12$. All training and generation tests were performed with TensorFlow 2.15.0 on a platform equipped with a Ryzen AI 9 HX 375 processor and an NVIDIA RTX 5090m GPU. The full training procedure is summarized in Algorithm~\ref{alg:training}.

\begin{algorithm}[t]
\caption{Training Stochastic Path Sampler}
\label{alg:training}
\begin{algorithmic}[1]
\STATE \textbf{Input:} Prior $\pi_0=\mathcal{N}(0,I)$, discretization steps $T$, batch size $N_s$, learning rate $\alpha$, training iterations $N$
\STATE \textbf{Output:} Trained SPS model
\FOR{$j=1$ to $N$}
    \STATE Sample mini-batch $\{\mathbf{s}_0^{(n)}\}_{n=1}^{N_s}\sim\pi_0$
    \STATE Set $\log q_\theta \gets \log\pi_0(\mathbf{s}_0)$
    \FOR{$i=0$ to $T-1$}
        \STATE Evaluate $K_{\theta,\mathrm F}(\mathbf{s}_i,t_i)$ and $\sigma_\theta(t_i)$
        \STATE Generate $\mathbf{s}_{i+1}$ using Eq.~\eqref{eq:training_forward_update} 
        
        \STATE Evaluate $K_{\theta,\mathrm B}(\mathbf{s}_{i+1},t_i)$

        \STATE Accumulate $\log q_\theta$ using  Eq.~\eqref{eq:training_logq}
    \ENDFOR
    \STATE Compute $\widehat{\mathcal{L}}_{\mathrm{SPS}}$ using Eq.~\eqref{eq:sps_minibatch_loss}
    \STATE Update network parameters by gradient descent
\ENDFOR
\end{algorithmic}
\end{algorithm}

\subsection{Generation and Independence Metropolis--Hastings Correction}
\label{app:generation}

After training, the learned forward dynamics is used as a proposal generator. Starting from $\mathbf{s}_0\sim\pi_0$, we integrate the same stochastic update in Eq.~\eqref{eq:training_forward_update} and obtain the endpoint $\mathbf{s}_T$, which is
identified with the proposed field configuration $\phi'$. The proposal log density $\log q(\phi')$ is evaluated along the generated path using
Eq.~\eqref{eq:training_logq}. The generation procedure is summarized in Algorithm~\ref{alg:generation}.

The generated proposal is then corrected by an Independence Metropolis--Hastings step.
For a current configuration $\phi$ and a proposed configuration $\phi'$, the acceptance
probability is
\begin{equation}
    A(\phi\rightarrow\phi')
    =\min\left[1,
    \frac{\tilde{\pi}^{*}(\phi')q_\theta(\phi;\tau)}{\tilde{\pi}^{*}(\phi)q_\theta(\phi';\tau')}
    \right], 
    \label{eq:imh_acceptance_appendix}
\end{equation}
where $\tilde{\pi}^{*}(\phi)=\exp[-S(\phi)]$ is the unnormalized target density. Since the proposal density $q_\theta(\phi;\tau)$ is evaluated along the generated trajectory (cf.~Eq.~\eqref{eq:tb_ratio_general}), the acceptance step is defined on the extended space of trajectories, on which detailed balance holds exactly; this is the extended-space Independence Metropolis--Hastings step referred to in the main text. This correction removes the residual bias caused by finite network expressivity and time discretization.

During generation, we use a finer time discretization than during training in order to reduce the discretization error of the stochastic trajectory. Specifically, the diffusion interval $[0,1]$ is discretized with $T=250$ steps during training and $T=2500$ steps during generation.

\begin{algorithm}[t]
\caption{Generation with Stochastic Path Sampler}
\label{alg:generation}
\begin{algorithmic}[1]
\STATE \textbf{Input:} Prior $\pi_0=\mathcal{N}(0,I)$, trained SPS model, generation steps $T$, batch size $N_s$
\STATE \textbf{Output:} Proposed configuration $\phi'=\mathbf{s}_{T}$ and proposal log density $\log q(\phi')$
\STATE Sample mini-batch $\{\mathbf{s}_0^{(n)}\}_{n=1}^{N_s}\sim\pi_0$
\STATE Set $\log q \gets \log\pi_0(\mathbf{s}_0)$
\FOR{$i=0$ to $T-1$}
    \STATE Evaluate the learned drift and diffusion networks
    \STATE Generate $\mathbf{s}_{i+1}$ using Eq.~\eqref{eq:training_forward_update} 
    \STATE Accumulate $\log q$ using Eq.~\eqref{eq:training_logq}
\ENDFOR
\STATE Return $\phi'=\mathbf{s}_T$ and $\log q_\theta(\phi';\tau')=\log q$
\end{algorithmic}
\end{algorithm}

\section{Magnetization Histogram}
\label{app:C}

The magnetization histograms shown in this appendix are obtained from the SPS proposals before the IMH correction. They demonstrate that, near the pseudocritical region and in the broken phase, the SPS proposal already populates both \(Z_2\)-related magnetization sectors, indicating that the learned \(Z_2\)-equivariant dynamics does not collapse onto a single mode. 

\begin{figure}[htbp]
\centering
\subfloat[$\kappa=0.28$]{
\label{fig:fig:magnet_kappa0.28}
\includegraphics[height=5cm,width=5cm]{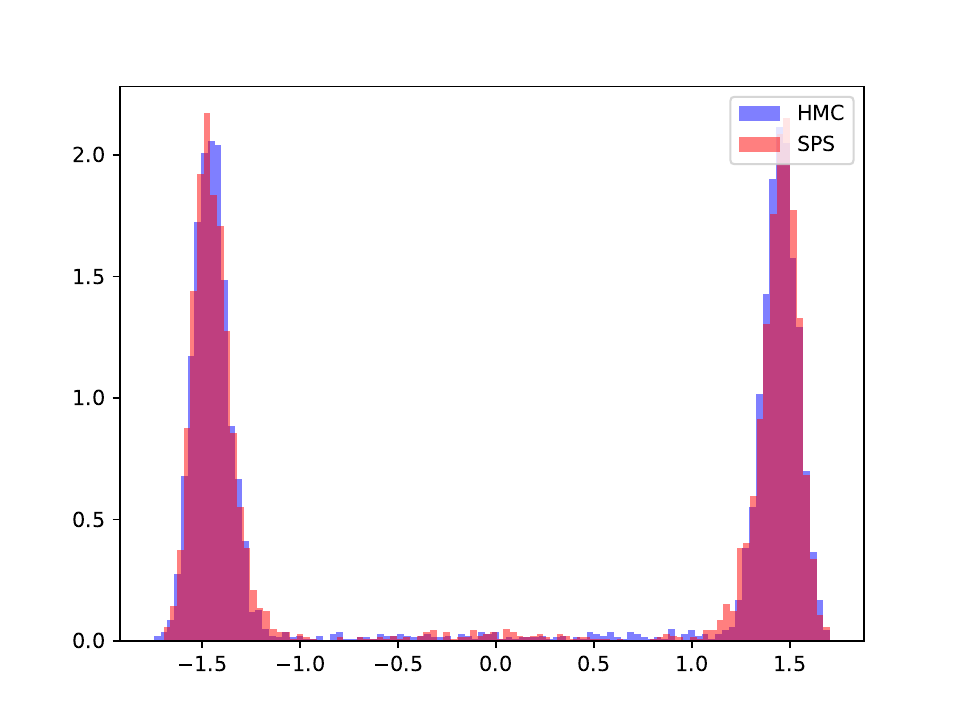}}
\subfloat[$\kappa=0.29$]{
\label{fig:fig:magnet_kappa0.29}
\includegraphics[height=5cm,width=5cm]{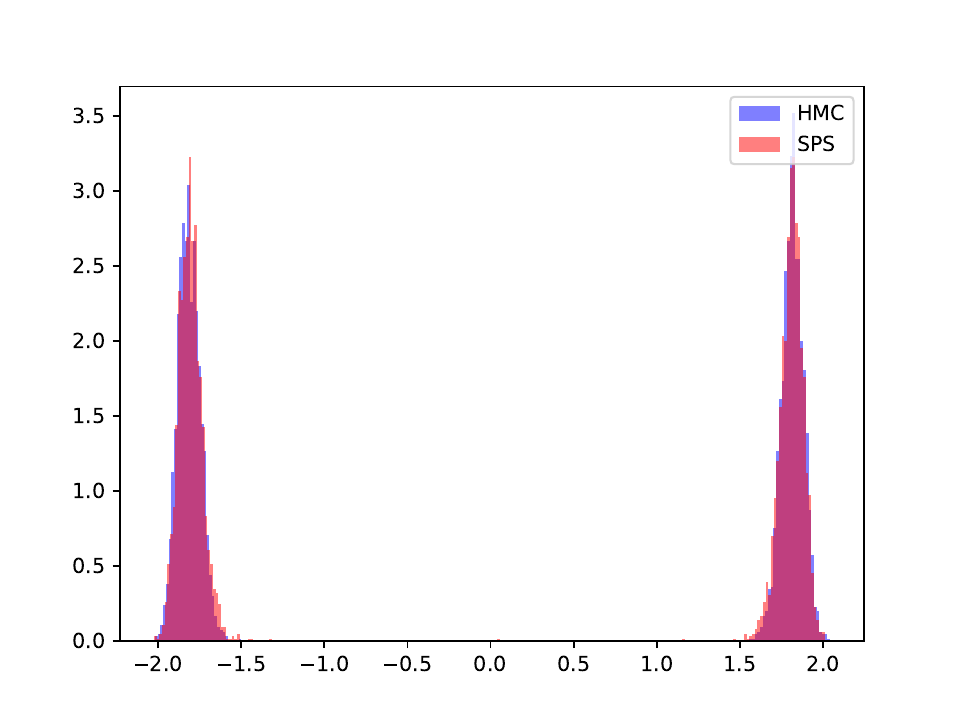}}
\subfloat[$\kappa=0.30$]{
\label{fig:magnet_kappa0.3}
\includegraphics[height=5cm,width=5cm]{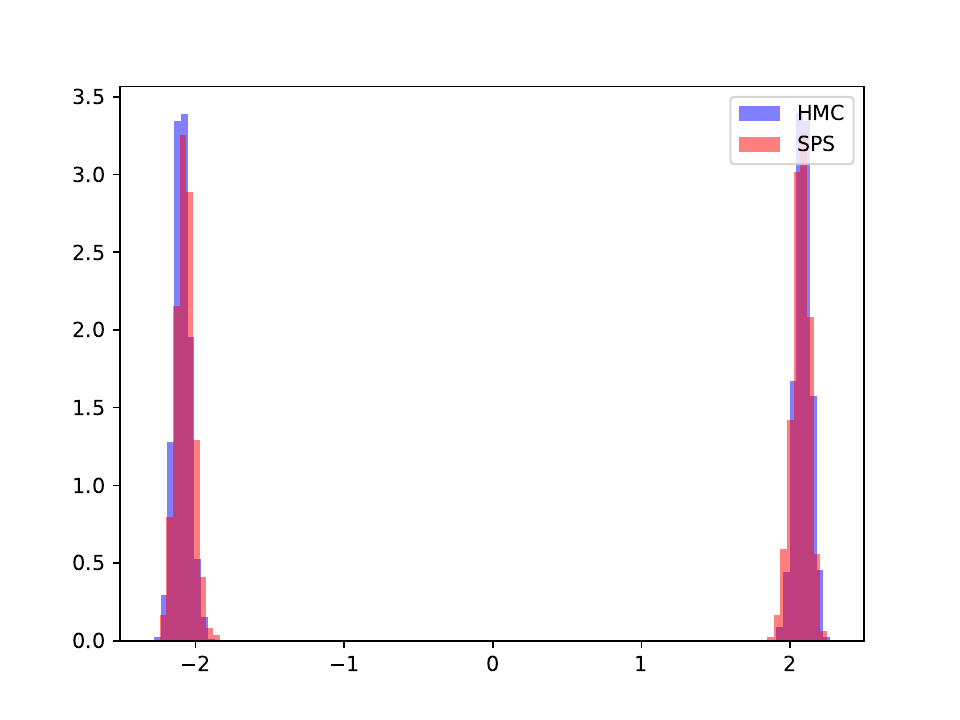}}
\caption{
Magnetization histograms at \(\kappa=0.28\), \(0.29\) and \(0.30\) for \(L=64\).  The blue histograms are obtained from HMC, while the red histograms are obtained from uncorrected SPS proposals before the IMH correction.  The two populated peaks correspond to the two \(Z_2\)-related magnetization sectors, indicating that the SPS proposal covers both modes in the broken phase. Note that the HMC and SPS histograms overlap almost completely, so that the blue (HMC) histogram is largely hidden beneath the red (SPS) one; the two distributions are statistically compatible.
}
\label{fig:Broken_Histogram}
\end{figure}

\begin{figure}[htbp]
\centering
\subfloat[$\kappa=0.25$]{
\label{fig:fig:magnet_kappa0.25}
\includegraphics[height=5cm,width=5cm]{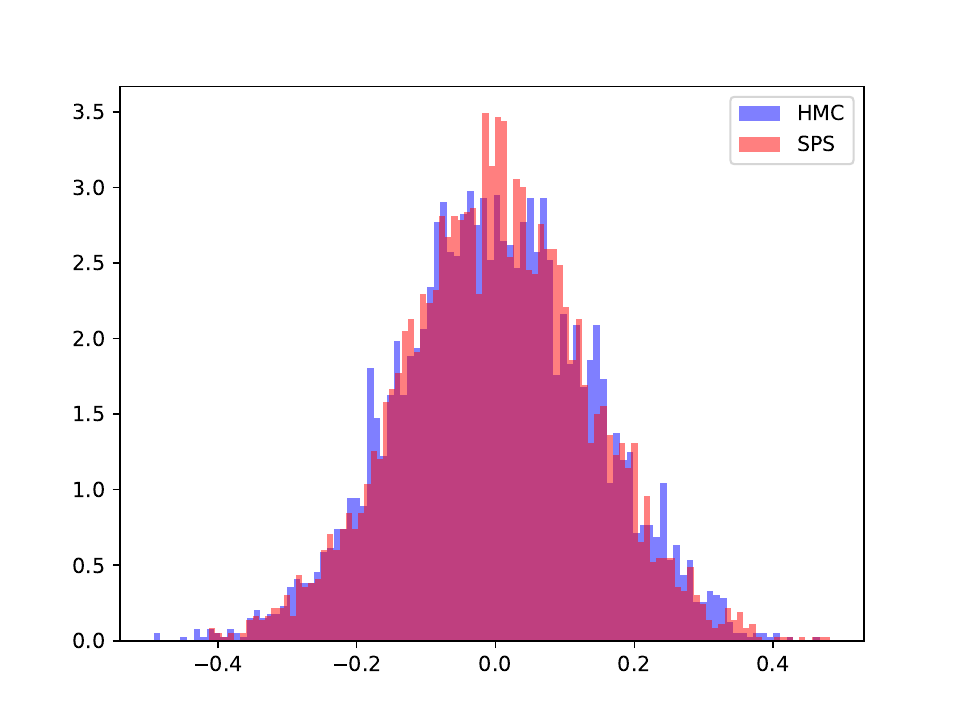}}
\subfloat[$\kappa=0.26$]{
\label{fig:fig:magnet_kappa0.26}
\includegraphics[height=5cm,width=5cm]{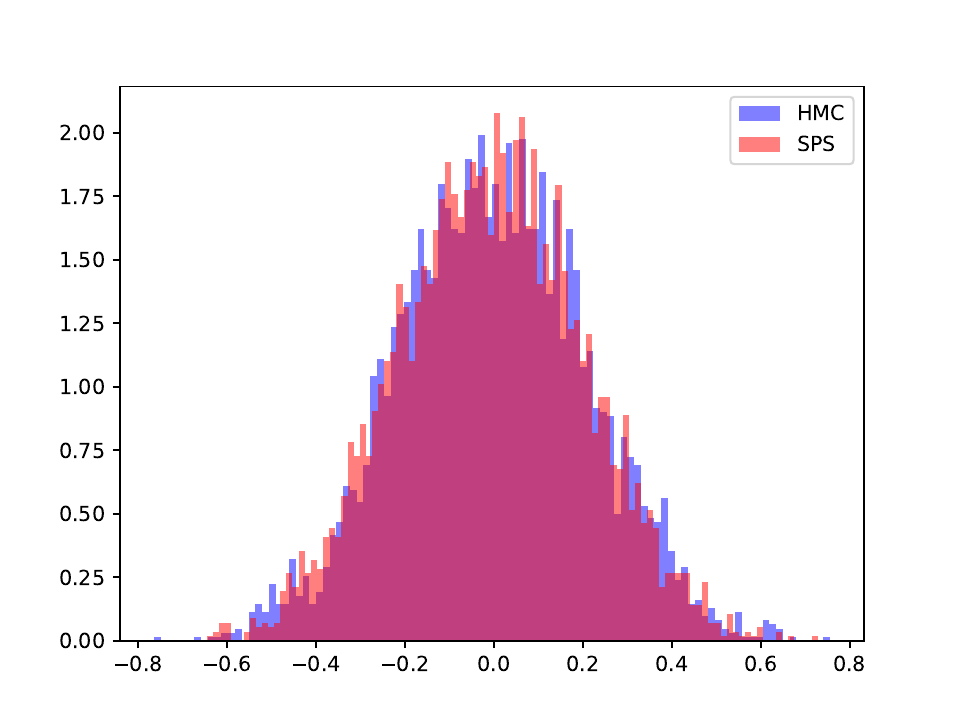}}
\subfloat[$\kappa=0.27$]{
\label{fig:magnet_kappa0.6}
\includegraphics[height=5cm,width=5cm]{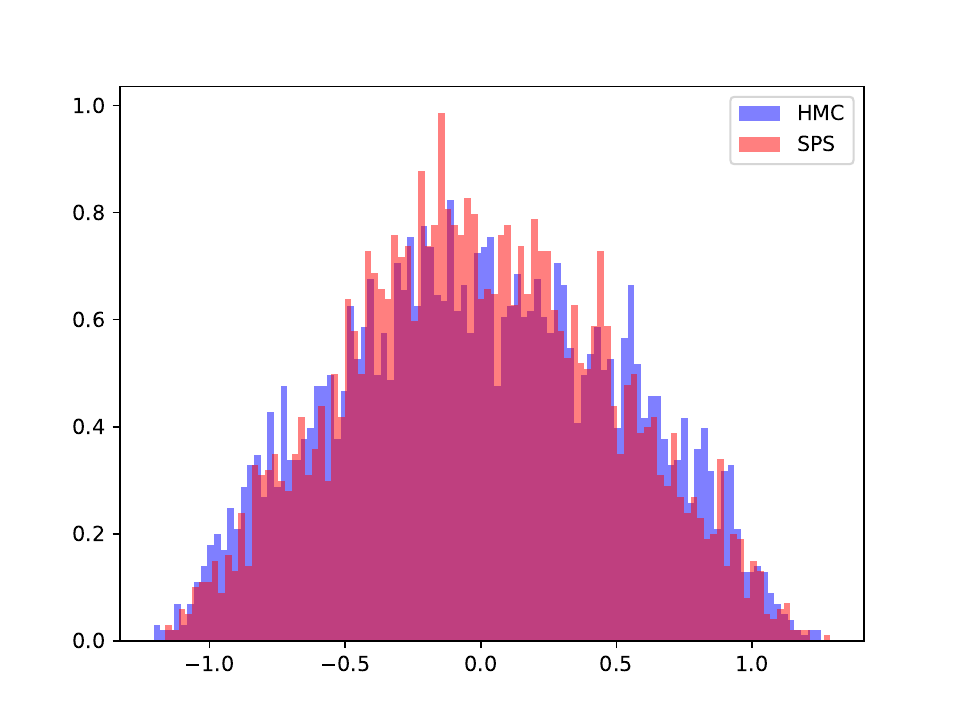}}
\caption{
Magnetization histograms for \(L=64\) at \(\kappa=0.25\), \(0.26\), and \(0.27\).  The blue histograms are obtained from HMC, while the red histograms are obtained from uncorrected SPS proposals before the IMH correction.  The agreement between the two distributions shows that the SPS proposal captures the broadening of the magnetization distribution as the system approaches the pseudocritical region. As in Fig.~\ref{fig:Broken_Histogram}, the two histograms overlap almost completely.
}
\label{fig:symmetric_Histogram}
\end{figure}
 
\section{Effective Sample Size and Acceptance Rate}
\label{app:ess_generation_depth}

\subsection{Effective Sample Size}

To investigate the dependence of the sampling efficiency on the lattice volume and the discretization depth used during generation, we evaluate the normalized effective sample size. For notational simplicity, we henceforth denote the trajectory-dependent importance weight $w(\phi;\tau)$ by $w(\tau)$, or simply by $w$ when no ambiguity arises. The normalised effective sample size is defined as
\begin{equation}
    \frac{\mathrm{ESS}}{N_{\mathrm{s}}}
    =
    \frac{
        \left(\sum_{i=1}^{N_{\mathrm{s}}} w_i\right)^2
    }{
        N_{\mathrm{s}}
        \sum_{i=1}^{N_{\mathrm{s}}} w_i^2
    },
\end{equation}
where \(w_i\) denotes the importance weight associated with the \(i\)-th generated trajectory. The normalized ESS lies in the interval \([1/N_{\mathrm{s}},1]\), with values close to unity indicating nearly uniform trajectory weights.

\begin{figure}[t]
\centering
\includegraphics[width=0.8\linewidth]{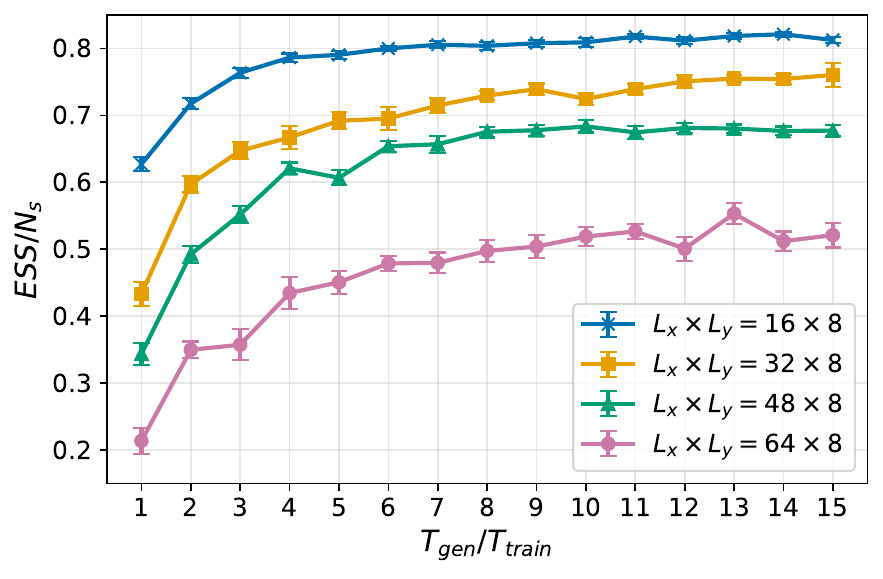}
    \caption{Normalized effective sample size as a function of \(T_{\mathrm{gen}}/T_{\mathrm{train}}\) at $\kappa=0.27$. The lines with star (blue), square (orange), triangle (green), and circle (purple) markers correspond to the \(16\times8\), \(32\times8\), \(48\times8\), and \(64\times8\) lattices, respectively. }
    \label{fig:ess_vs_generation_depth}
\end{figure}

Fig.~\ref{fig:ess_vs_generation_depth} shows the normalized ESS at $\kappa=0.27$ as a function of the ratio $T_{\mathrm{gen}}/T_{\mathrm{train}}$, where \(T_{\mathrm{train}}=250\) is the number of stochastic steps used during training and \(T_{\mathrm{gen}}\) is the number of steps used during generation. 

For all lattice sizes, increasing \(T_{\mathrm{gen}}\) initially produces a rapid improvement in the ESS. This shows that trajectories generated with the same discretization depth as that used during training still contain a non-negligible finite-step error, which can be reduced by evaluating the learned dynamics with a time discretization finer than the one used during training. The improvement saturates at large \(T_{\mathrm{gen}}/T_{\mathrm{train}}\), where a volume-dependent ESS is reached.


In Fig.~\ref{fig:V_std}, we demonstrated that the variance of the logarithmic trajectory weights scales with the volume, i.e.,
\begin{equation}
    \operatorname{Var}(\log w)\propto |\Lambda|.
    \label{eq:Var_volume}
\end{equation}
If we assume that these logarithmic trajectory weights are approximately Gaussian,
\begin{equation}
   X \equiv \log w \sim \mathcal{N}(\mu,\sigma^2),
    \label{eq:logw_normal}
\end{equation}
we can approximate the normalized ESS in the large-sample limit as
\begin{equation}
    \frac{\mathrm{ESS}}{N_{\mathrm{s}}}
    \simeq
    \frac{\langle w\rangle^2}{\langle w^2\rangle}.
\end{equation}
Since for log-normal weights,
$\langle w\rangle=e^{\mu+\sigma^2/2}$ and $\langle w^2\rangle=e^{2\mu+2\sigma^2}$,
we then find
\begin{equation}
    \frac{\mathrm{ESS}}{N_{\mathrm{s}}}
    \simeq
    e^{-\sigma^2}
    =
    e^{-\operatorname{Var}(\log w)}.
    \label{eq:ess_normal}
\end{equation}
The normalized ESS is therefore expected to decrease exponentially with the lattice volume. Allowing for finite-volume and finite-generation-depth effects, we fit the data to
\begin{equation}
    \frac{\mathrm{ESS}}{N_{\mathrm{s}}}
    \simeq A e^{-a|\Lambda|}.
\end{equation}
Using the results at the largest generation depth considered, we obtain \(A\simeq0.98\) and \(a\simeq1.1\times10^{-3}\), showing that the observed volume dependence is consistent with an approximately exponential decrease over the range of lattice sizes studied here.

\begin{figure}[htbp]
\centering
\includegraphics[width=0.8\linewidth]{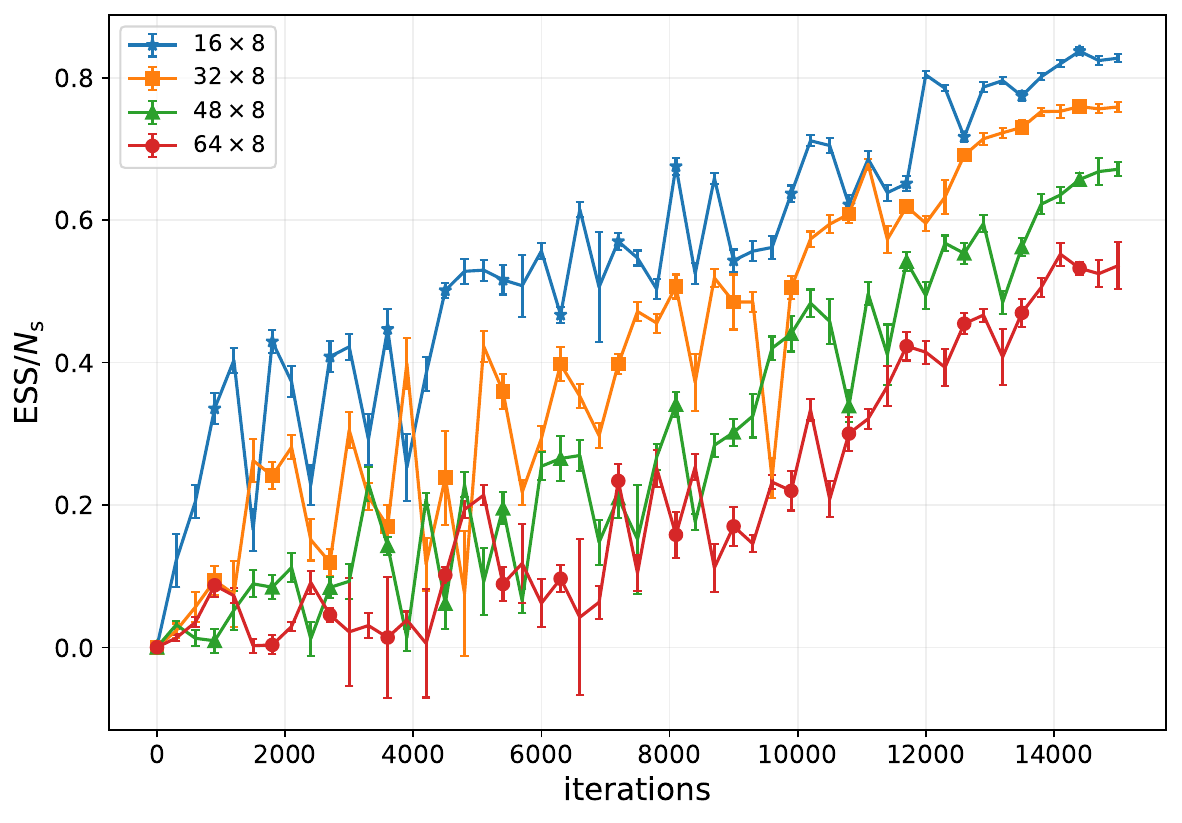}
\caption{
Normalized effective sample size as a function of training iterations for lattices \(16\times8\), \(32\times8\), \(48\times8\), and \(64\times8\) at \(\kappa=0.27\), with \(T_{\mathrm{gen}}/T_{\mathrm{train}}=10\) and \(N_s=4096\). The lines with star, square, triangle, and circle markers correspond to the \(16\times8\), \(32\times8\), \(48\times8\), and \(64\times8\) lattices, respectively.}
\label{fig:ess_vs_epoch}
\end{figure}

Estimates of the scaling with the volume depend on the network architecture and the details of the training. To demonstrate this, we show in Fig.~\ref{fig:ess_vs_epoch} the dependence of the ESS on the number of epochs at a fixed architecture. We observe that the ESS increases with the number of training iterations. 
To quantify the cost required to maintain a fixed sampling efficiency, we may define \(N_{\rm iteration}^{*}(|\Lambda|)\) as the number of training iterations required to reach a prescribed normalized ESS. As shown in Fig.~\ref{fig:ess_vs_epoch}, the number of training iterations required to reach the same normalized ESS increases systematically with the lattice volume at fixed \(T_{\mathrm{gen}}/T_{\mathrm{train}}=10\).

\subsection{Relation between the ESS and Acceptance Rate}

The volume dependence of the normalized ESS also has a direct implication for the scaling of the acceptance rate. Here we provide a derivation of the ESS-based prediction for the acceptance rate introduced in Eq.~\eqref{eq:ess_based_prediction}. We first derive the relation between the acceptance rate and the variance of the logarithmic importance weights under the Gaussian approximation \eqref{eq:logw_normal}. Combining this result with the relation between the normalized ESS and the log-weight variance \eqref{eq:ess_normal} then yields the ESS-based acceptance-rate prediction.  The derivation is performed directly on the extended trajectory space, so that the path dependence of $q_\theta(\phi;\tau)$ at finite training is retained.

Let $\tau$ denote the current trajectory and $\tau'$ an independently generated proposal trajectory. At stationarity, the current trajectory is distributed according to the reweighted forward path measure, whereas $\tau'$ is drawn independently from $dq_{\mathrm F}$. Using the importance weight defined in Eq.~\eqref{eq:reweight_ensemble}, the stationary mean acceptance rate corresponding to Eq.~(B.21) is
\begin{align}
\langle A\rangle
&=
\frac{1}{\mathbb E_{dq_{\mathrm F}}[w]}
\mathbb E_{dq_{\mathrm F}(\tau)dq_{\mathrm F}(\tau')}
\left[
w(\tau)
\min\left(
1,
\frac{w(\tau')}{w(\tau)}
\right)
\right]
\nonumber\\
&=
\frac{
\mathbb E_{dq_{\mathrm F}(\tau)dq_{\mathrm F}(\tau')}
\left[
\min\!\left(w(\tau),w(\tau')\right)
\right]
}{
\mathbb E_{dq_{\mathrm F}}[w]
}.
\label{eq:appendix_exact_acceptance}
\end{align}
In the second line, we used
\begin{equation}
w(\tau)
\min\left(
1,\frac{w(\tau')}{w(\tau)}
\right)
=
\min\!\left(w(\tau),w(\tau')\right).
\label{eq:min_weight_identity}
\end{equation}
Eq.~\eqref{eq:appendix_exact_acceptance} is the independent Metropolis--Hastings representation of the stationary mean acceptance rate in terms of importance weights~\cite{Liu1996,Tierney98,Billera01}. No assumption on the shape of the weight distribution has been introduced up to this point.

We now use the Gaussian approximation for the logarithmic weights introduced in Eq.~\eqref{eq:logw_normal}, and let $X'$ denote an independent copy of $X$ associated with the proposed trajectory. By symmetry,
\begin{equation}
\mathbb E[\min(w,w')]
=
2\,\mathbb E
\left[
e^X\,\mathbb I(X\leq X')
\right].
\label{eq:min_weight_symmetry}
\end{equation}
Multiplication of the Gaussian density of $X$ by $e^X$ gives
\begin{equation}
e^x p_X(x)
=
\mathbb E[w]\,
\frac{1}{\sqrt{2\pi\sigma^2}}
\exp\left[
-\frac{(x-\mu-\sigma^2)^2}{2\sigma^2}
\right].
\label{eq:gaussian_tilt}
\end{equation}
Therefore, the logarithmic weight associated with the reweighted current trajectory follows $\mathcal N(\mu+\sigma^2,\sigma^2)$, whereas the logarithmic weight of the independently generated proposal follows $\mathcal N(\mu,\sigma^2)$.

Let $X_\ast$ denote the logarithmic weight of the reweighted current trajectory. Using Eq.~\eqref{eq:min_weight_symmetry}, Eq.~\eqref{eq:appendix_exact_acceptance} can then be written as
\begin{equation}
\langle A\rangle
=
2\,
\mathbb P\!\left(
X_\ast\leq X'
\right),
\label{eq:acceptance_probability_gaussian}
\end{equation}
with
\begin{equation}
X_\ast
\sim
\mathcal N(\mu+\sigma^2,\sigma^2).
\label{eq:tilted_logweight_distribution}
\end{equation}
Since $X_\ast$ and $X'$ are independent, Eq.~\eqref{eq:tilted_logweight_distribution} implies
\begin{equation}
X'-X_\ast
\sim
\mathcal N(-\sigma^2,2\sigma^2).
\label{eq:logweight_difference_distribution}
\end{equation}
It follows from Eq.~\eqref{eq:acceptance_probability_gaussian} and Eq.~\eqref{eq:logweight_difference_distribution} that
\begin{align}
\langle A\rangle
=
2\,\mathbb P(X'-X_\ast\geq0)
=
2\Phi\left(
-\frac{\sigma}{\sqrt{2}}
\right)
=
\operatorname{erfc}\left(
\frac{\sigma}{2}
\right),
\label{eq:appendix_acceptance_sigma}
\end{align}
where $\Phi$ denotes the cumulative distribution function of the standard normal distribution.

Using the ESS--variance relation in Eq.~\eqref{eq:ess_normal}, we obtain
\begin{equation}
\sigma^2
=
-\ln\left(
\frac{\mathrm{ESS}}{N_{\mathrm s}}
\right).
\label{eq:sigma_from_ess}
\end{equation}
Substituting Eq.~\eqref{eq:sigma_from_ess} into Eq.~\eqref{eq:appendix_acceptance_sigma} yields
\begin{equation}
\langle A\rangle
\simeq
\operatorname{erfc}\left[
\frac{1}{2}
\sqrt{
-\ln\left(
\frac{\mathrm{ESS}}{N_{\mathrm s}}
\right)
}
\right].
\label{eq:appendix_ess_acceptance}
\end{equation}
Thus, under the log-normal approximation for the SPS importance weights, both the normalized ESS and the mean acceptance rate are controlled by the same quantity, $\sigma^2=\operatorname{Var}(\log w)$. Eq.~\eqref{eq:appendix_ess_acceptance} contains no additional fit parameters; its accuracy is determined by how well the finite-sample weight distribution is described by the log-normal approximation.
This result is used in the main part of the paper in Fig.~\ref{fig:Acceptance}.

\subsection{Volume Scaling of the Acceptance Rate}

The relation derived above also determines the volume dependence of the IMH acceptance rate. As shown in Eq.~\eqref{eq:Var_volume}, the variance of the logarithmic trajectory weights is approximately extensive. Writing $\sigma^2\simeq a|\Lambda|$ and substituting this scaling into Eq.~\eqref{eq:appendix_acceptance_sigma} gives
\begin{equation}
\langle A\rangle
\simeq
\operatorname{erfc}
\left(
\frac{1}{2}\sqrt{a|\Lambda|}
\right).
\label{eq:acceptance_volume_scaling}
\end{equation}
For moderate log-weight fluctuations, $a|\Lambda|\lesssim 1$. Expanding the erfc function as $\operatorname{erfc}(x) = 1-2x/\sqrt{\pi}+\mathcal O(x^3)$ gives
\begin{equation}
1-\langle A\rangle
\simeq
\sqrt{\frac{a|\Lambda|}{\pi}}.
\label{eq:acceptance_moderate_volume}
\end{equation}
In the same regime, expanding Eq.~\eqref{eq:ess_normal} gives $1-\mathrm{ESS}/N_{\mathrm s}\simeq a|\Lambda|$. Thus, although the normalized ESS exhibits a deficit that grows linearly with the lattice volume, the IMH acceptance deficit in Eq.~\eqref{eq:acceptance_moderate_volume} grows only as the square root of the volume. This explains why the ESS can exhibit a visible volume dependence while the acceptance rate remains comparatively high over the lattice sizes considered here.

For completeness, in the asymptotic large-volume regime, where $a|\Lambda|\gg 1$, applying
\begin{equation}
\operatorname{erfc}(x)
\simeq
\frac{e^{-x^2}}{\sqrt{\pi}\,x}
\label{eq:erfc_asymptotic}
\end{equation}
to Eq.~\eqref{eq:acceptance_volume_scaling} gives
\begin{equation}
\langle A\rangle
\simeq
\sqrt{
\frac{4}{\pi a|\Lambda|}
}
\exp\left(
-\frac{a|\Lambda|}{4}
\right).
\label{eq:acceptance_asymptotic_volume}
\end{equation}
Hence, at sufficiently large volume the IMH acceptance rate also decreases exponentially with $|\Lambda|$, with an exponential rate four times smaller than that of the normalized ESS in Eq.~\eqref{eq:ess_normal}.


\providecommand{\href}[2]{#2}\begingroup\raggedright\endgroup

\end{document}